\documentclass[8.5pt,preprint]{revtex4}
\usepackage{balance} 

\usepackage{graphicx} 
\usepackage{lastpage}
\usepackage[format=plain,justification=raggedright,singlelinecheck=false,font=small,labelfont=bf,labelsep=space]{caption} 
\usepackage{fancyhdr}

\usepackage{float}
\usepackage{booktabs}

\usepackage{latexsym} 
\usepackage{times} 
\usepackage{graphicx}
\usepackage[tight,FIGTOPCAP]{subfigure}
\usepackage{color}
\usepackage{afterpage}
\usepackage{multirow}
\usepackage[latin1]{inputenc}
\usepackage[english]{babel}

\begin{document}

\title{Gene clusters reflecting macrodomain structure respond to nucleoid perturbations$^\dag$}

\author{\textbf{Vittore F. Scolari,\textit{$^{a, b, c}$} Bruno Bassetti,\textit{$^{c, d}$} Bianca Sclavi,\textit{$^{e}$} and Marco Cosentino Lagomarsino$^{\ast}$\textit{$^{a, b, c}$}}}\vspace{0.5cm}

\begin{abstract}
Focusing on the DNA-bridging nucleoid proteins
  Fis and H-NS, and integrating several independent experimental and
  bioinformatic data sources, we investigate the links between
  chromosomal spatial organization and global transcriptional
  regulation. By means of a novel multi-scale spatial aggregation
  analysis, we uncover the existence of contiguous clusters of
  nucleoid-perturbation sensitive genes along the genome, whose
  expression is affected by a combination of topological DNA state and
  nucleoid-shaping protein occupancy. The clusters correlate well with
  the macrodomain structure of the genome. The most significant of
  them lay symmetrically at the edges of the \emph{ter} macrodomain
  and involve all of the flagellar and chemotaxis machinery, in
  addition to key regulators of biofilm formation, suggesting that the
  regulation of the physical state of the chromosome by the nucleoid
  proteins plays an important role in coordinating the transcriptional
  response leading to the switch between a motile and a biofilm
  lifestyle.
\end{abstract}

\newcommand{\bra}[1]{\langle #1 |}
\newcommand{\ket}[1]{| #1 \rangle}
\newcommand{\braket}[2]{\langle #1 | #2 \rangle}
\newcommand{\fig}[1]{\ref{fig:#1}}
\newcommand{\separator}{\begin{center}\rule{\textwidth}{1px}\end{center}}
\newcommand{\todo}[1]{\textcolor{red}{#1}}
\newcommand{\rf}[1]{\todo{[ref#1]}}
\newcommand{\de}{\mathrm{d}}

\heavyrulewidth=.08em
\lightrulewidth=.05em
\cmidrulewidth=.03em
\belowrulesep=.65ex
\aboverulesep=.4ex
\cmidrulesep=\doublerulesep
\cmidrulekern=.25em
\defaultaddspace=.5em


\let\thefootnote\relax\footnotetext{\dag~Electronic Supplementary Information (ESI) available: on website \url{http://www.lgm.upmc.fr/scolarietal/}.
}


\footnotetext{\textit{$^{a}$~Genomic Physics Group, FRE 3214 CNRS "Microorganism
  Genomics"}}
\footnotetext{\textit{$^{b}$~Universit\'e Pierre et Marie Curie, 15 rue de L'\'Ecole de M\'edecine Paris, France}}
\footnotetext{\textit{$^{c}$~Universit\`a degli Studi di Milano, Dip.  Fisica.  Via Celoria 16, 20133 Milano, Italy}}
\footnotetext{\textit{$^{d}$~I.N.F.N. Milano, Via Celoria 16, 20133 Milano, Italy}}
\footnotetext{\textit{$^{e}$~LBPA, UMR 8113, ENS Cachan, 61, avenue du President Wilson, 94235 Cachan, France}} 
\footnotetext{$^{\ast}$~[ e-mail address:]{Marco.Cosentino-Lagomarsino@upmc.fr}}


\maketitle

\section*{Introduction}

The success of a cell's survival under different growth conditions
depends on its ability to regulate a coordinated transcriptional
response to specific environmental changes or stresses, involving
large groups of genes~\cite{ABL+03,CdL03}.  In bacteria,
transcriptional regulation of genes depends on the binding of proteins
to DNA, but also on the physical configurations of the resulting
mesoscopic protein/DNA complex, the
nucleoid~\cite{Dillon2010a,TM05,JC06}.

Specific nucleoid-shaping transcription factors (NSTFs) have the task
of modulating the nucleoid's dynamic structure in response to changes
in environmental conditions~\cite{LNW+06}. This can result, for
example, in a change in the compaction of the chromosome and in a
differential distribution of mechanical energy, stored as
supercoiling. These changes in the physical properties of the DNA can
affect the level of expression of specific genes, in parallel to the
activity of specific transcription factors. NSTFs may thus change the
expression of many genes (some of which may code for the same NSTFs)
both directly and through the physical conformations that they induce
on the genome~\cite{Dillon2010a}.

The current transcription network view of gene regulation represents
specific transcription-factor binding sites upstream of promoters as a
directed graph, linking each transcription factor to its target node
(which represents the transcript and its protein product) if the
transcription factor has at least one binding site with documented
activity in the cis-regulatory region of the
target~\cite{BLA+04}. With this definition, the interaction graph
structure is given by both large-scale experiments and collections of
small-scale experiments~\cite{GJP+08}.
This view considers the graph of all genome (transcriptional)
interactions but completely disregards the effects on gene expression
due to changes of the nucleoid.
The role played by the nucleoid's structure in the hierarchy of events
leading to large scale transcription patterns remains largely to be
elucidated.  In the near future it will be necessary to incorporate
these into a generalized description of the organization of the genome
and the regulatory network it encodes for. This is a challenging
problem because the output of the transcription network is due to the
sum of both local and global regulatory signals, from the biochemical
properties of transcription factors, such as their concentration and
affinity for the sites on the promoter, to the mesoscopic nucleoid
organization and DNA conformational and topological states.

Nucleoid organization itself remains to be fully characterized. Most
of NSTFs have been identified~\cite{LNW+06}, and in some cases their
local action is well known: for example, DNA bridging by H-NS or Fis is
believed to be important for DNA loop
stabilization~\cite{DLK+05}. Such supercoiled plectonemic loops can be
topologically isolated from the rest of the genome, modulating protein
binding and transcription rates~\cite{TM05}.
On larger scales, meticulous recombination experiments have shown that
the nucleoid as a polymer-protein complex is divided into six
compartments, or ``macrodomains''~\cite{VPR+04}. Four of these
macrodomains are defined by preferential interaction of DNA fragments
within the same domain and by their spatial colocalization within the
cell~\cite{EVE+07,boccard1,VPR+04}.
DNA segments within a macrodomain typically intersect, while
inter-macrodomain collisions appear to be restricted. The two
remaining ``nonstructured regions'' are more fluid in nature.
Other experiments have probed large-scale nucleoid
structure by tagging of specific loci and/or by nucleoid isolation
~\cite{EMB08,wiggins,WRS08,WLP+06,CWO05,COS+01,Zim06b-a}. GFP-RNAP
fusions have also been used in this context in an effort to identify
possible ``transcription factories''~\cite{CJ06,JC06,LDC+05,CJ03}.

The relation between nucleoid state and gene expression has been the
subject of several recent studies.  Experimentally, this question has
been addressed using mainly transcriptomics and chromatin
immunoprecipitation combined with microarrays (ChIP-chip).
Thanks to elegant experiments linking the length distribution of
observed supercoiled domains with the transcriptional response to
locally induced supercoil relaxation~\cite{PHA+04}, we know that gene
expression can be affected by its localization within such an isolated
topological domain~\cite{cozzarelli,MRB05}. Thus, the same bridging nucleoid
proteins that give the nucleoid a branched structure when observed
by electron microscopy may also be responsible for part of the
the transcriptional regulation not accounted for by the network of
transcription factors and regulated genes.
Other experiments have characterized RNA-polymerase (RNAP) and specific NSTF
binding by ChIP-chip~\cite{WSB+07,grainger,cho2}.  By
monitoring generic protein binding throughout the
chromosome~\cite{vora}, Vora~\emph{et~Al.} have found extended
protein binding regions (called ``extended protein
occupancy domains'' or EPODs) connected to either transcriptionally
silent or highly expressed clusters of genes. 
Finally, transcriptomics has been applied to the study of the effects
of the knockout 
of specific bridging nucleoid protein and to changes in the level of
negative supercoiling~\cite{blot,MGH+08,bradley}.

Computationally, attempts have been made to characterize the binding
specificity of NSTFs~\cite{LBB+07,Hengen1997} and to interrogate the
one-dimensional arrangement of genes of different categories for
signals possibly related to nucleoid
structure~\cite{WtW04,Sonnenschein2009}. In particular, a thread of
studies~\cite{Mathelier2010,WKC+07,Kep04,kepes1} on
possible ``periodicity'' patterns has found a number of interesting
spatial regularities in the arrangement of genes belonging to
different categories. For example, the correlation of gene expression
with gene codon bias, has recently been related to large contiguous
``sectors'' along the chromosome~\cite{Mathelier2010}.

Here, we present an integrated analysis combining different
independent data sources that report on (i) specific DNA-protein
interactions, (ii) different levels of gene expression, and (iii)
large scale nucleoid structure, in order to uncover coherent,
consistent correlations between these different levels of genome
organization.
We focus on the spatial distribution along the genome of these data
sets.  Our statistical aggregation analysis shows that part of the
macrodomain structure of the genome emerges directly from the analysis
of the distribution of genes that change their expression when
comparing wild-type versus nucleoid-perturbed
conditions~\cite{MGH+08}. In addition, by the analysis of specific
nucleoid protein occupancy profiles and EPODS, we recover a similar
structure, where the most significant regions flank the Ter
macrodomain of the nucleoid defined by Valens and
coworkers~\cite{VPR+04}. In the same data, we also recover a
periodicity signal in analogy with previous studies on other genomic
data sources related to transcription of \emph{E.~coli} genes and
genome adaptation. 
Finally, a functional analysis of the clusters reveals significant
enrichment of the flagellar and chemotaxis genes, together with
regulators of biofilm formation.

\section{Results}

\subsection{The distribution of genes sensitive to H-NS/Fis deletion and
  supercoiling perturbations is nonuniform along the genome.}

We began the analysis from the transcriptomics data of
Blot~\emph{et~al.} \cite{blot,MGH+08}. In these experiments the global
expression profiles of wild-type \emph{E.~coli} K12 were compared with
mutants carrying combined nucleoid perturbations in the form of
knockouts of the NSTFs proteins Fis or H-NS, and a mutation of a gyrase
or a topoisomerase, affecting the average supercoiling background.
The differential analysis of these experiments gives
seven sets of genes significantly responding to these
nucleoid-related perturbations (Supplementary Figure \ref{fig:liste}).
A simple density map of these lists shows notable peaks that
correspond to linear regions of the chromosome characterized by a
stronger transcriptional reaction to nucleoid-related perturbations
with respect to other parts of the chromosome.

\subsection{Clusters of transcriptional response to H-NS/Fis deletion
  and supercoiling perturbations correlate with macro-domain and
  chromosomal segment organization of the genome. }

\begin{figure}[h]
  \centering
  \includegraphics[width=.6\columnwidth]{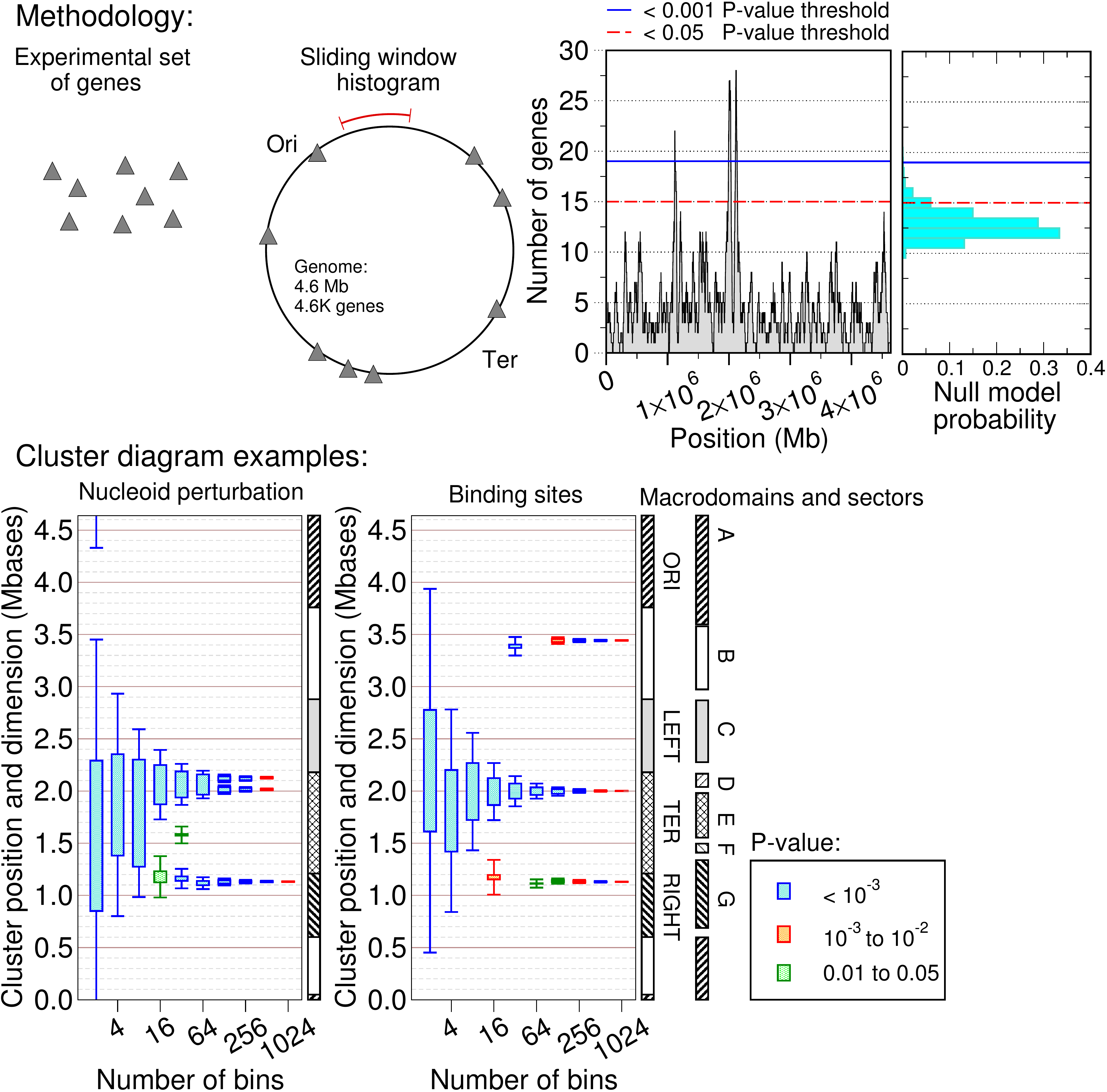}
  \caption{Procedure followed to detect one-dimensional aggregation of
    genes in the lists. Top panel: A sliding-window density histogram
    associates to every coordinate on the circular genome the number
    of genes in the empirical list in an interval surrounding the
    point and spanning a fixed bin-size. As an example, the density
    histogram of the WT-$\Delta$H-NS(low $-\sigma$) list at bin-size
    $b_s = L/256$ is shown; the density of each position is compared
    with the $P$-value thresholds from the null model (Methods) in
    order to obtain the significant positions, which are in turn
    merged with a compatibility threshold of size $b_s$ in order to
    define the clusters. Bottom panel: example of cluster diagram for
    the WT-$\Delta$H-NS(low $-\sigma$) transcriptomics data, and the
    FISbind(Grainger) protein binding data. The $y-$axis is the genome
    coordinate. The plot shows the clusters discovered at different
    scales as a function of the number of bins $L/b_s$ ($x-$axis). The
    box indicates the position of the peak while the whiskers span the
    maximal extension of the clusters. Both cluster diagrams give
    clusters that localize close to the edges of the Ter macrodomain
    at small scales ($L/1024 \simeq 5Kb$ to $L/16 \simeq 0.3 Mb$), and
    cover the Ter or the Ter+Left region at larger scales ($L/4 \simeq
    1.2Mb$ or more). The binding data have a cluster in the
    nonstructured Left region appearing only at small scales ($L/16$
    or less).}
\label{fig:histclust}
\end{figure}

Marr and coworkers~\cite{MGH+08} detected clusters in the same gene
lists with a threshold technique linking genes with the proximity
threshold $t$ in the range $1b < t < 10 Kb$.  We performed a different
clustering analysis probing multiple scales, considering also the
statistical significance of the one dimensional aggregation by
comparing the peaks of the empirical histogram with the highest peak
found in the histogram of randomized lists. Our method can thus define
relevant clusters and also associate a $P$-value to them for every given
scale (see Methods and Figure \ref{fig:histclust}).

We systematically analyzed both supercoiling sensitive genes in the
WT, $\Delta$Fis and $\Delta$H-NS intra-strain lists (see Methods) and
the nucleoid-protein sensitive genes at fixed supercoiling condition
in the WT-$\Delta$Fis(low $-\sigma$), WT-$\Delta$Fis(high $-\sigma$),
WT-$\Delta$H-NS(low $-\sigma$), WT-$\Delta$H-NS(high $-\sigma$)
inter-strain lists. Supplementary Figure \ref{fig:cluexpr} shows the
clusters found. 
A cluster at 1.90-1.96 Mb, or 41 centisomes ($P = 0.01$) of genes
sensitive to supercoiling changes was found for all three
strains. Another cluster at 1.10-1.20 Mb, 24 centisomes, ($P <
10^{-3}$) sensitive to supercoiling changes appears in the wild type
experiment, and a cluster at 3.80-3.81 Mb, 82 centisomes ($P = 0.04$)
is present in the mutant lacking the Fis protein. The deletion of Fis
at fixed supercoiling background causes a variation in gene expression
in a cluster at 1.95-2.03 Mb, 43 centisomes, ($P = 0.05$) in both high
and low negative supercoiling conditions and in a cluster at 3.33-3.55
Mb, 74 centisomes ($P = 0.01$) only at low negative supercoiling
conditions. The deletion of H-NS at fixed supercoiling background
causes the emergence of a pair of neighboring clusters respectively at
1.97-2.05 and 2.08-2.15 Mb (around 43 centisomes) and a cluster at
1.06-1.17 Mb (24 centisomes) at both supercoiling conditions ($P <
10^{-3}$). Another cluster emerges in the condition of low negative
supercoiling at 1.50-1.66 Mb (34-35 centisomes, $P = 0.03$), a
position compatible with the replication terminus, and a cluster at
0.30-0.34 Mb, 7 centisomes, ($P = 0.05$) appears at high negative
supercoiling condition.
Notably, very similar clusters appear from the data on the genes
significantly responding to Fis deletion as a function of the growth
phase, specifically in mid exponential phase~\cite{bradley}
(Supplementary Figure \ref{fig:blubrad}).

\begin{figure}[t]
  \includegraphics[width=.5\columnwidth]{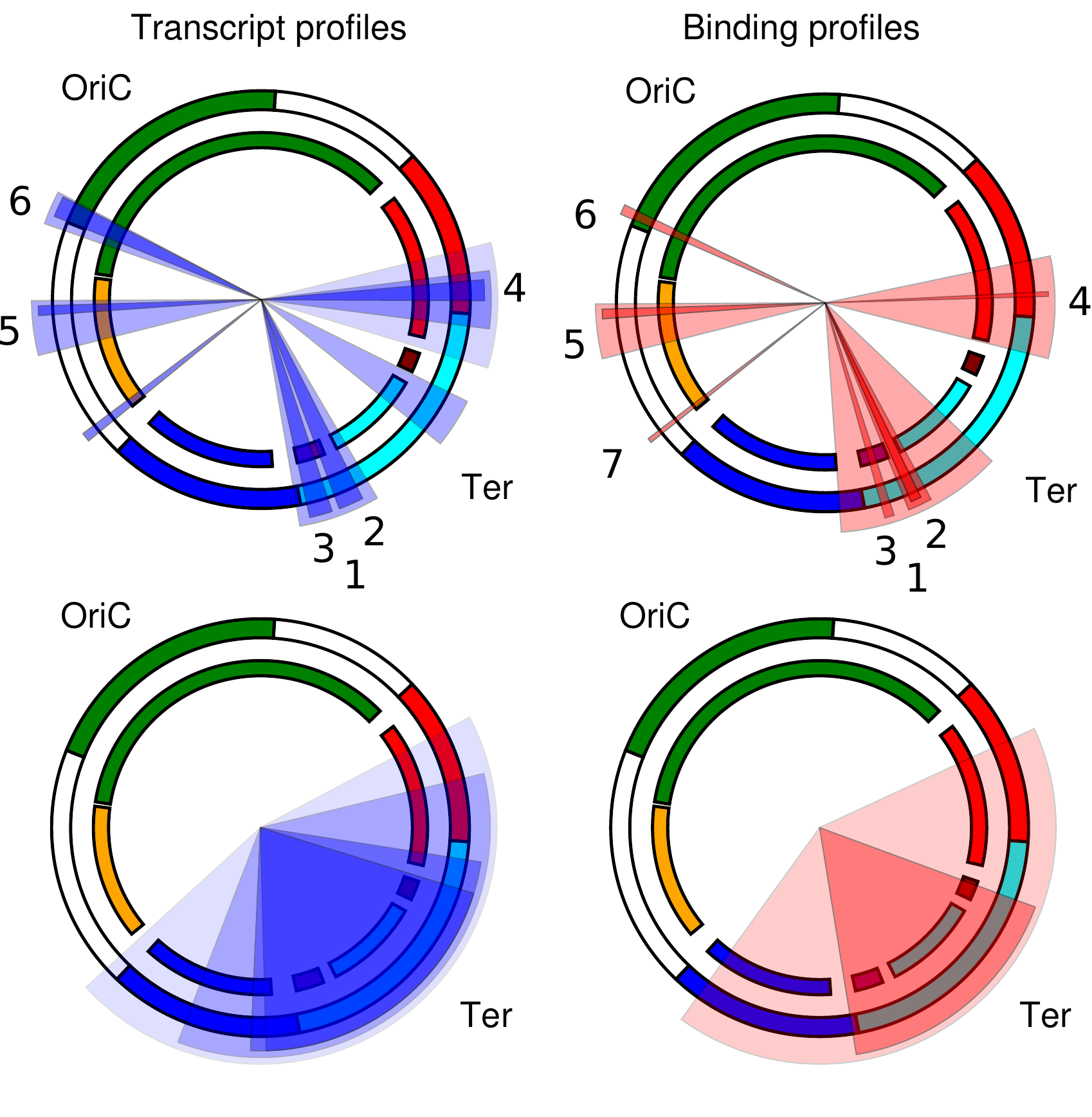}
  \caption{Graphical representation of the most significant clusters
    found for nucleoid perturbation transcriptomics data (left) and
    for data of direct binding of Fis and H-NS (right). The clusters
    found are represented by colored wedges with transparency
    increasing with size.  The clusters found at large observation
    scales ($L/8 \simeq 0.6Mb$ or more) are shown separately in the
    lower panels.  In the drawings, the outer colored circle
    represents macrodomains while inner colored circle contains the
    chromosome sectors defined by Mathelier and Carbone
    \cite{Mathelier2010}. The numbering of the small clusters is
    described in Supplementary Table \ref{tab:space}. Note the
    compatibility of the clusters found with the edges of the Ter and
    Ori macrodomains at small scales and with different segments of
    the Ter region at larger scales.}
\label{fig:summa}
\end{figure}

A summary of the most significant clusters is shown in Figure
\ref{fig:summa} and Supplementary Table \ref{tab:space}.  At the scale
of the macrodomains, these clusters appear to overlap well with the
previously observed segmented structure of the
chromosome~\cite{VPR+04,Mathelier2010}. At smaller scales, clusters
preferentially localize towards the edges of macrodomains or in
nonstructured regions (Supplementary Figures \ref{fig:cluexpr} to
\ref{fig:clubind}).
In particular, the highest significance clusters appear at all scales
in H-NS-related perturbation experiments as well as in response to
supercoiling changes in the intra-strain WT list (see Figure
\ref{fig:summa}).  At small scales, these clusters concentrate rather
well, within the uncertainty of the analysis, to the edges of the Ter
macrodomain defined by Boccard and coworkers \cite{VPR+04,EMB08}, also
termed fluid region by a more recent study~\cite{wiggins}, while at
larger scales they cover the entire Ter macrodomain.

\subsection{Binding profiles of nucleoid-shaping proteins along the
  genome follow the same spatial patterns of the expression data.}

The previous analysis shows a tight link of the transcription program
with the structuring of the nucleoid, and the macrodomains in
particular, related to the action of the NSTFs Fis and H-NS. In order
to test for the direct action of these proteins at these sites, we
considered data concerning specific binding sites of Fis and H-NS on
the chromosome.
Binding sites of H-NS and Fis proteins along the \emph{E.~coli} genome
\emph{in vivo} were extracted from the ChIP-chip data of
Grainger~\emph{et~al.}~\cite{grainger}. We also considered the list of
FIS binding sites obtained by high-density ChIP-chip by
Cho~\emph{et~al.}~\cite{cho2} and from the RegulonDB
database~\cite{GJP+08} (see methods for the nomenclature of these
experiments).

The cluster diagrams of those lists are reported in Supplementary
Figure \ref{fig:clubind}. 
The clearest clusters appear from the experiments of Grainger and
coworkers.  In particular, we found three clusters with a low
$P$-value ($P < 10^{-3}$) in FISbind(Grainger) of which two (at
1.12-1.14 Mb, 24 centisomes, and at 1.95-2.03 Mb, 43 centisomes) that
are precisely superimposed on the clusters found in Wt-$\Delta$H-NS
gene expression data sets, laying at the border of the Ter
macrodomain, and one (at 3.43-3.46 Mb, 74 centisomes) corresponding to
the cluster already found in WT-$\Delta$Fis. We also found two
clusters in HNSbind(Grainger) at 1.99-2.02 Mb, 43 centisomes ($P <
10^{-3}$) and at 2.09-2.12 Mb, 45 centisomes ($P = 0.01$) that
correspond to the cluster found in the transcriptomics in
Wt-$\Delta$H-NS set, at the border of the Ter macrodomain. Finally, we
find a cluster at 2.98-3.00, 64.5 centisomes ($P = 0.01$) and another
one at 3.78-3.82, 82 centisomes ($P < 10^{-3}$) that can be
superimposed with the border of the Ori macrodomain. The Fis binding
data of Cho \emph{et al.} appear to give a weaker signal for this
cluster, which however remains visible.
This discrepancy may come from the different resolution of the two
studies, or from a different threshold used in the two studies to
define a binding site.
Finally, the other control set from H-NS high-resolution binding data
from Oshima \emph{et al.}  shows all the clusters found in the
Grainger data, plus additional clusters, especially found in the Ter
area. In this case, the differences are most likely due to the
different growth conditions in this experiment.

Considering Fis and H-NS binding sites from the RegulonDB dataset, we
found multiple clusters of difficult interpretation. This is possibly
due to the knowledge bias of those data and the elusive nature of NSTF
binding motifs~\cite{LBB+07}.
Sonnenschein and coworkers~\cite{Sonnenschein2009} found that genes
that, according to current knowledge as compiled by RegulonDB, do not
participate in the transcriptional regulatory network, show a
statistical repulsion for linear genome positioning with genes that
do, and they interpret this as a signature of nucleoid-related gene
regulation.
Motivated by this finding, we also performed a comparative cluster
analysis on the genes included in the RegulonDB regulatory network and
outside (Supplementary Figure \ref{fig:netdiagram}).
While the knowledge-biased RegulonDB network presents many clusters of
difficult interpretation, we surprisingly found that the genes outside
the RegulonDB network show clear clusters on the border of the Ter
macrodomain and a larger scale cluster covering the Ter macrodomain
($P < 10^{-3}$). These regions might be enriched of genes expressed
under physical control by the nucleoid rather than by specific network
interactions. Alternatively, the unknown function of some of these
operons results in a lack of information regarding their regulatory
mechanisms.

In order to gain more insight into the binding of nucleoid related
proteins on the DNA, we also considered the extended protein occupancy
domains (EPOD) experimental data produced by Vora \emph{et al}
\cite{vora}. This technology reveals protein occupancy across an
entire bacterial chromosome at the resolution of individual binding
sites.  EPODs are long contiguous segments of DNA-bound proteins along
the chromosome, and were found to correspond to transcriptionally
silent region (tsEPODs) or highly expressed ones (heEPODs). 
We compared the density profiles of heEPOD and tsEPOD to the gene
density of the other data sets. The local contributions to the Pearson
correlation coefficient of some representative data sets are plotted
in Supplementary Figure \ref{fig:epodcorr}. The figure shows a high
correlation between heEPOD density, FISbind(Grainger) and
WT-$\Delta$H-NS(low $-\sigma$), specifically in the area around the
border of the Ter macrodomain (around $2 Mb$, or 43 centisomes). The
similarity of the density profile is evident from the visual
inspection of the normalized densities. As an example, Figure
\ref{fig:epodsuper} reports the case of heEPOD and WT-$\Delta$H-NS(low
$-\sigma$).
Finally, we considered the enrichment of lists of genes within EPODs
with respect to genes responding to nucleoid perturbations
(Supplementary table \ref{tab:overlap_he_zero} and
\ref{tab:overlap_ts_zero}). heEPODs have a significant intersection
with all the inter-strain experiments as well as with the intra-strain
supercoiling changes in absence of H-NS.  On the other hand, tsEPODs
significantly overlap only with the perturbation experiments related
to H-NS (inter-strain), as expected from its known role as a
transcriptional repressor \cite{Ball1992, Dorman2004}.

\subsection{Functional classes of genes in the clusters.}

Hypergeometric testing of enrichment for MultiFun~\cite{Serres2000}
functional categories was carried out systematically to all the
considered datasets and also to the genes contained in the most
significant clusters found in the spatial aggregation analysis.
The result of the enrichment analysis are available on the web site
\url{http://www.lgm.upmc.fr/scolarietal/}.

The flagella and chemotaxis classes, in addition to several biofilm
related classes, are enriched in both of the clusters at the edges of
the Ter macrodomain (Cluster 1 and Cluster 4). These classes are also
enriched in the following datasets: FISbind (Grainger), HNSbind
(Grainger) $\Delta$Fis intra-strain, WT-$\Delta$Fis and WT-$\Delta$H-NS
at both supercoiling conditions, WT-$\Delta$H-NS (ME), and
WT-$\Delta$Fis (150 and 240min) in the Bradley data sets.
In order to compare these clusters the flagella synthesis network, we
carried out a spatial aggregation analysis of the operons directly
controlled by the FlhC transcription factor, the master regulator of
flagella gene expression. The cluster diagram is reported in
Supplementary Figure \ref{fig:flaclust} and shows a clear correlation
with the clusters identified by the analysis of the datasets from both
Fis and H-NS binding and transcriptomics experiments.

The role of Fis and H-NS in the regulation of flagellar gene
expression has been known for some time \cite{Bertin1994,
  Kelly2004,Ko2000,Soutourina2003}.  Interestingly, flagella and
chemotaxis genes share the same clusters with functional classes
related to biofilm formation, such as the operons responsible for
curli and capsule synthesis and the M and O antigens, in addition to
phospholipid synthesis. The genes in each experimental list can be in
turn divided into two sets, depending on whether their level of
expression increases or decreases upon a given perturbation. These
sets are shown in Supplementary Table~\ref{tab:tablex}. The genes for
motility and those for biofilm formation are found for the most part
on opposite sets. As a consequence, a given perturbation will affect
the properties of most of the genes within this cluster by a change of
gene expression in opposite directions for genes in the the two
functional categories.

\subsection{Statistically significant periodicities emerge in the
  arrangement of nucleoid-perturbation sensitive genes along the
  genome.}
While the analysis described above identified specific large regions
of high density of affected genes, previous studies have identified
specific regular structures of the nucleoid that suggest a more
global regulatory influence of nucleoid
organization~\cite{Kep04,JAK04,WKC+07,Mathelier2010}. We thus 
wanted to test the possibility that the genes that are sensitive to 
supercoiling variation and to deletion of Fis and H-NS
may be organized in regularly spaced groups on the chromosome. We
built the histogram of the position of the genes and the histogram of
the distance between each pair of genes in the empirical lists of the 
chromosome.

The height of the spectral peaks for every periodicity in the
empirical distributions was compared to the distribution of global
maxima of a random null model shuffling the gene lists in order to
discern statistically significant periodicities (see Methods and
Supplementary Figures \ref{fig:figperi} and \ref{fig:figperi_pos}). 

Supplementary Table \ref{tab:period} contains a synthetic summary of
the periodicities found.  A significant periodicity was found in the
position distribution of the intra-strain WT list with period length
of 352 Kb and a $P$-value $< 0.04$; a similar periodicity was found
also in the distance distribution with a period of 328 Kb ($P< 0.01$).
Another significant periodicity emerges in the $\Delta$Fis data set,
with period length of 101 Kb ($P< 0.01$); this periodicity was also
confirmed by a similar signal in the distance distribution at about 98
Kb.
The $\Delta$H-NS set shows a periodicity ($P< 0.05$) appearing in the
distance distribution at 20 Kb (which is below the resolution of the
density histogram).
In the WT-$\Delta$H-NS(low $-\sigma$) two periodicities emerge at 385
and 660 Kb ($P< 0.01$). Both of them were confirmed in the distance
distribution with a period of 331 and 589 Kb ($P<0.01$). A highly
significant peak also emerges at the very large scale of 2.3 Mb, which
is a sign of a genome wide asymmetry that confirms the results of the
cluster analysis. A periodicity of 100 Kb also exists only in the
distance distribution ($P< 0.05$).

In the WT-$\Delta$H-NS(high $-\sigma$) list, two periodicities were
found at 385 and 675 Kb ($P< 0.01$). Both of them were
confirmed in the distance distribution with periodicity of 370 and 660
Kb  ($P< 0.01$). Two periodicities at 22Kb and 100Kb were
also found only in the distance distribution ($P< 0.05$),
while on the position distribution there is a non significant local
maximum at 100Kb. No significant periodicity was detected in either
the WT-$\Delta$Fis(low $-\sigma$) and WT-$\Delta$Fis(high $-\sigma$)
empirical lists.
It is worthwhile noticing that the periods detected in the distance
distribution are systematically smaller than periods detected in the
density. However, the discrepancy is relatively small ($< 10\%$) and
always smaller than the bin-size of the density histogram.

In synthesis, we found the following significant periodicities,
reported in supplementary table \ref{tab:period}: one of $20 \pm
36$Kb, for $\Delta$H-NS and WT-$\Delta$H-NS (high -$\sigma$), that could
be comparable to the length of supercoiling domains~\cite{PHA+04}, and
one at $100 \pm 36$Kb, the same length of the solenoidal signal found
by Wright and Kepes~\cite{Kep04}.
In addition, a periodicity around $360 \pm 36$ Kb, which might be
related to macrodomain size, appears under perturbation by both H-NS
deletion and changes in supercoiling. The compatibility threshold of
$36$ Kb was set to twice the bin-size of the density distribution.

\section*{Discussion}


\paragraph*{Efficient detection of linear aggregation.  } The large
amount of high-throughput data being generated regarding genome
organization and transcription requires the development of efficient
approaches that are able to integrate different data sources.
We developed a novel strategy to quantify the linear aggregation along
the genome of different gene lists. This technique has the two main
advantages of considering linear aggregation at all scales, and of
being able to assign statistical confidence to the presence of gene
clusters at different scales by comparing empirical data with suitable
null models. We applied this method to multiple sets of experiments
related to nucleoid protein binding and to the global transcriptional
response to nucleoid perturbation, with a focus on the two
nucleoid-shaping proteins Fis and H-NS.

\paragraph*{Clusters of contiguous genes responding transcriptionally
  to nucleoid perturbations appear to follow the macrodomain structure
  of the genome.  }
The results of the analysis confirm that the transcriptional response
to nucleoid perturbations in the form of Fis/H-NS deletion and changes
in the average level of supercoiling is highly non uniform along the
genome~\cite{MGH+08}, and reflected the results obtained from the
analysis of binding profiles (see below).  All the data analyzed,
coming from independent sources~\cite{blot,MGH+08,bradley}, show
multiple significant clusters whose arrangement is highly correlated
with the probed spatial structure of the chromosome, and specifically
with
macrodomains~\cite{VPR+04,MRB05,boccard1,EB06,wiggins}. Macrodomains
have a well-defined spatial arrangement and localization in the cell
both during chromosome segregation and during interphase, and preserve
the linear order of genes along the
genome~\cite{EMB08,EVE+07,Roc08,EB06,WLP+06,NLY+06,RWS08}.
At larger scales, generally the clusters appear to overlap well with
macrodomains, while at smaller scales they preferentially localize
towards the edges of macrodomains or in nonstructured regions. This is
particularly evident for the data producing clusters in the Ter
region.
These clusters also superimpose well with the segments of coherence
between gene expression and codon bias~\cite{Mathelier2010}.  Thus, we
conclude that this evidence supports a tight link between large-scale
transcription programs of the cell and the spatial organization of the
genome as a nucleoprotein-polymer complex~\cite{Muskhelishvili2010}.

Specifically, Fis and H-NS are known to be important protein factors
for the shaping of the nucleoid~\cite{Dillon2010a,LNW+06}. Fis is
believed to create and isolate supercoiling domains by bridging two
distant DNA regions, and is generally associated with positive
transcriptional control.  H-NS forms stable oligomers that bridge two
DNA helices, stabilizing a plectoneme, and thus possibly inhibiting
transcription~\cite{DLK+05,Fang2008}. It is worthwhile noting that in
the nucleoid perturbation experiments considered here, the clusters
emerging from changing the supercoiling background and those that
emerge under H-NS deletion are very similar, which suggests that the
effects of these two different perturbations on the nucleoid might be
related.
A similar clustering behavior has recently been reported for the
transcriptional response to deletions of the nucleoid protein
Hu~\cite{Berger2010}.
In this study it was found that the frequency of genes influenced by
the absence of Hu correlates with the macrodomain organization of the
nucleoid. The genes upregulated in the absence of Hu are found for the
most part in the Ori macrodomain extending to the nearby nonstructured
regions, while the genes downregulated in these conditions are found
with higher frequency in the Ter and Right macrodomains.  In addition
the pattern of transcription upregulation in the Hu mutant mirrors the
density of gyrase sites along the chromosome, pointing to a specific
role of Hu in maintaining the supercoiling homeostasis in the
rRNA-rich Ori macrodomain. On the other hand, the genes in the Ter
macrodomain appear to be more sensitive to global changes in
supercoiling and to regulation by Fis and H-NS in mid exponential phase
(when rRNA expression is excluded).
However, the Hu regulon does not show significant overlap with the Fis
and H-NS regulons and with the genes influenced by
supercoiling~\cite{blot}, only a small subset of genes are found to be
co-regulated~\cite{Oberto09}.

\paragraph*{Clusters of nucleoid-shaping protein binding spatially
  correlate with nucleoid perturbations.  }

The same clustering analysis applied to the binding of Fis and H-NS
from ChIP-chip data~\cite{grainger,cho2,Oshima2006} reveals clusters
that correspond very well with those emerging from the transcriptomics
data sets. We can also report that our preliminary survey of very recent
ChIP-seq data for the same proteins also agrees with these
findings~\cite{Seshasayee2010}.
This confirms that the two proteins also have a direct role in
physically shaping the region of the nucleoid that responds to their
action. More specifically, deletion of H-NS influences the expression
of the genes in the cluster at 1.1Mb (24 centisomes), which overlaps
with the cluster of Fis binding from the ChIP-Chip dataset, pointing
to a tight link between the activity of these two proteins.
%
%
Moreover the cluster at 2Mb (43 centisomes) is also enriched for
heEPODS, high expression extended protein occupancy
domains~\cite{vora}.

It is important to note that while the two transcriptomics datasets
and the EPOD results were obtained from cells growing in rich media
(LB or YT), the ChIP-chip data from the other two experiments was
obtained from cells growing in minimal media (M9 plus glucose or
fructose), with the exception of the Oshima data, wich refer to
mid-exponential growth phase in LB medium.
Most data relate to cells in mid exponential phase, which suggests
that the coherent results between binding patterns and transcriptional
response to perturbation might correspond to growth-phase specific
features.
In addition, some variation may also arise from the loose definition
of ``mid exponential phase'', determining when the cells are harvested
after dilution in fresh medium. The agreement of these different
datasets indicates that the positioning of the clusters is robust
within this range of experimental conditions. The dataset that
addresses directly the changes in gene expression as a function of
growth phase, shows that the clusters reflect changes in cellular
metabolism (see Supplementary Figure~\ref{fig:blubrad}). In the
future, we expect that further studies will directly probe the role of
nucleoid structure and organization in response to changes of both
growth rate and growth phase.

\paragraph*{Coherent periodicity signals emerge from both nucleoid
  protein binding and transcriptional response to nucleoid
  perturbations .  }
We also found significant evidence for periodic arrangement of the
nucleoid perturbation sensitive genes.  Some of these periodicities
correspond to characteristic lengths that can be associated to
supercoil domains, the $\sim$20Kb branched structure of genomic
plectonemes~\cite{PHA+04} or the previously observed 100Kb periodicity
of evolutionarily conserved gene sets~\cite{WKC+07}.  Notably, as in
the clustering analysis, the periodicities emerging from supercoil
perturbation also correspond to those related to H-NS deletion.
Evidence for spatial organization of genes along the chromosome has
already been presented for \emph{E. coli K12}
\cite{Kep04,JAK04,WKC+07}, and models that could explain it have been
formulated in the form of the so-called ``rosette model'' \cite{cook},
and the ``solenoid model'' \cite{kepes1,Kep04}.
In the case of our data the question remains open regarding whether
the periodic signal is simply related to the presence of
clusters. This is technically difficult to test, as it would require a
null model that randomizes a list by keeping its linear aggregation
properties constant. It is possible that newly developed techniques
are effective in bypassing this problem~\cite{Junier2010}.

\paragraph*{The two main clusters at the edges of the Ter macrodomain
  include the whole flagella regulon and key regulators of biofilm
  formation. }
We will now focus on the possible functional aspects of the clusters.
This analysis has identified two clusters of genes on either side of
the Ter macrodomain (Cluster 1 and 4) whose expression is affected by
deletion of either Fis or H-NS and upon changes in negative
supercoiling~\cite{MGH+08,blot,bradley}, in addition these clusters
superimpose with those obtained from the analysis of Fis and H-NS
binding obtained by ChIP-chip~\cite{grainger,cho}. The list of genes
in these two clusters include all the operons for flagellar proteins
and several genes required for cell adhesion and the formation of
biofilms. The expression of flagella is induced when the bacteria need
to swim either away from a stress or towards a richer nutrient
environment. Swimming is also necessary for the first steps of
biofilms formation leading to reversible attachment. However, flagella
synthesis is soon shut down as the biofilm structure begins to form
(reviewed in \cite{Pruess2006,Beloin2008a}). A similar exclusive gene
expression program takes place when the cells transit from exponential
growth into stationary phase. For example, below 30 $^o$C the genes
needed for the synthesis of curli fimbriae are induced, while
expression of the flagella is repressed via a regulatory network where
the second messenger c-di-GMP is important~\cite{Weber2006}.

Expression of flagella takes place in a sequential pattern
corresponding to the order of assembly of the different protein
components \cite{Chilcott2000,Kalir2001}. The flagellar expression
cascade is controlled by a master regulator FlhDC, and FliA, the
flagella-specific sigma factor. This regulatory network also includes
post-transcriptional and post-translational regulation and feedback
control~\cite{Kalir2005,Smith2009a}.

\paragraph*{Symmetric clustered organization of the flagella
  regulon. }
The organization of the flagellar regulon in the two nucleoid-related
clusters identified here could suggest a differential regulation of
these two sets of genes. However, the sequential order of expression
is not reflected in their linear organization along the chromosome
(Supplementary Table \ref{tab:tablex}). On the other hand, their
position on opposite sides of the two replicores suggests that it may
bring an advantage to replicate these two clusters roughly
simultaneously instead of sequentially, in order to maintain the
relative proportions of the flagellar proteins.

Moreover, the two sides of the regulon would attain on average equal
accumulated supercoiling from replication rounds, which would make (in
absence of stable topological barriers) symmetrically placed genes on
different replichores sense similar physical cues.  
This may confer
a specific sensitivity of these regions to changes in supercoiling and
nucleoid protein abundance that play a role in differential expression
under different kinds of stresses.

Another interesting question is whether this symmetry is common across
bacterial species. Certainly this is true for close species such as
\emph{Salmonella}. On the other hand, there are indications that this
symmetric arrangement might be a general principle. Studies of
bacterial comparative genomics~\cite{WKC+07} show a tendency of
cofunctional genes to be placed symmetrically with respect to
replication origins. Other studies~\cite{Darling2008,Eisen2000}
uncovered evidence for preferred symmetric chromosomal inversions
around the replication origin in evolving bacteria, which would
preserve a symmetric arrangement of genes.

\paragraph*{Are the flagella regulon clusters part of a
  hyperstructure? }
At the same time, it is plausible that the two clusters are in contact
or found near each other in the cytoplasm due to the compaction of the
Ter macrodomain~\cite{Mercier2008}. One can speculate that these
clusters are part of a larger structure that is co-localized in
three-dimensional space, and that both spatial aspects and
physico-chemical ones contribute to its function. Following the
example of the eukaryotic field, these spatial relationships could
emerge from both computational~\cite{janga1} and experimental
studies~\cite{Duan2010}.
The preference of a possible structure that colocalizes genes to be
mirror-symmetric can be argued by the fact that it would be disrupted
only once per replication round by advancing replication forks.

\begin{figure}[h]
{\center
  \includegraphics[width=0.4\columnwidth]{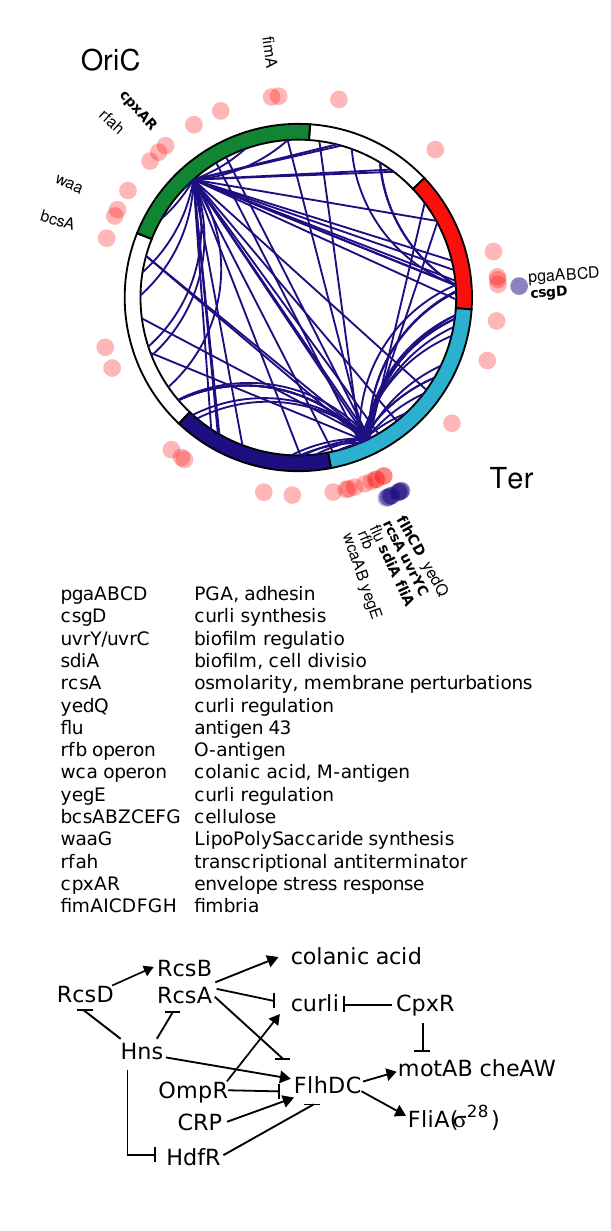}}
\caption{ (top panel) The main genes for flagella, chemotaxis and
  biofilm formation are partitioned between the Ori and the Ter
  macrodomains and particularly within Cluster 1 and 4 obtained from
  the analysis (Figure~\ref{fig:summa}).  The blue dots correspond to
  the flagellar and chemotaxis genes. The red dots correspond to the
  genes related to biofilm formation (increased color intensity
  results from presence of closely spaced genes).  The unlabeled red
  dots correspond to various cryptic fimbriae and adhesins operons,
  according to the lists in Supplementary Table~\ref{tab:tabley}. The
  gene names in bold correspond to transcription factors, and the blue
  lines to interactions from these transcription factors to their
  targets (RegulonDB). (bottom panel) Direct and indirect regulation
  of flagellar and curli expression, see
  refs. \cite{Beloin2008a,Hengge2009b,Ko2000,Krin2010,Ogasawara2010,Saldana2009,Wang2004,Korea2010,Prigent-Combaret2001}. For
  additional levels of regulation mediated by c-di-GMP see
  \cite{Pesavento2009,Hengge2009b,Weber2006}.}
\label{fig:figThree}
\end{figure}

\paragraph*{Direct and indirect nucleoid-based regulation of the
  flagella-regulon clusters. }
As described above, our analysis has identified these clusters because
of the high density in genes whose expression is affected by specific
nucleoid perturbations and/or high density of nucleoid protein binding
sites. Supercoiling, Fis and H-NS are known regulators of flagellar
expression in several organisms~\cite{Bertin1994, Kelly2004,
  Ko2000,Soutourina2003}. It is useful to discuss the effects that
these parameters may have on the expression of the global regulators
or directly on the different genes found downstream in the regulatory
network.

H-NS can influence the expression of the FlhDC transcription factor
both directly and indirectly through the regulation of expression of
the HdfR and RcsAB transcription factors \cite{Soutourina1999,
  Ko2000,Krin2010} resulting in a feed forward loop mode of regulation
(Figure \ref{fig:figThree}). RcsAB also represses the curli fimbriae
operon in Cluster 4 (csgDEFG, csgBAC) and positively regulates the
expression of the colanic acid operon in Cluster 3 (20 genes). Colanic
acid is an important component of the capsule necessary for mature
biofilm formation \cite{Beloin2008a}. The gene for RcsA is adjacent to
one of the flagellar regulon in Cluster 2 and is induced in the strain
lacking H-NS (Supplementary Table \ref{tab:tabley}), while the gene
for RcsB, which is also involved in the regulation of acid stress
response, is found in the Left macrodomain and is independently
regulated. As possible evidence for direct regulation by H-NS, a
cluster of H-NS binding is observed in the ChIP-chip data for Cluster
1 but not for Cluster 4 (Table \ref{tab:space}), nevertheless some
predicted H-NS sites are also found near or at three of the operons in
Cluster 4: csgDEFG, flgMN and flgBCDEFGHIJ (Supplementary Table
\ref{tab:tabley}) \cite{LBB+07}.  

This direct regulation by H-NS can in turn be differentiated into two
nonexclusive mechanisms of transcription regulation: H-NS can directly
influence the binding of RNA polymerase but it can also result in a
change in DNA topology by the formation of oligomeric structures
\cite{Fang2008}.  For example, the flagellar genes are in general
activated by the presence of H-NS independently of the supercoiling
state of the DNA, in part probably because of the induction of the
fliA gene, however the ones in Cluster 4, including the FlgM
anti-sigma factor, are also induced by a change in topology
(Supplementary Table \ref{tab:tabley}).
On the other hand, the flhD gene is found on all the hyp lists,
confirming that it is activated by an increase in negative
supercoiling, independently of the presence of Fis or H-NS, in
accordance with experimental \emph{in vitro} and \emph{in vivo}
observations \cite{Soutourina2002}.

As already mentioned for the regulation by RcsAB, in addition to the
flagellar operons, Cluster 1 and 4 also contain several other genes
involved in biofilm formation and maturation that are influenced by
the different perturbations to the nucleoid, as shown in Supplementary
Table \ref{tab:tabley}. These include the genes coding for the
O-antigen, the M-antigen, colanic acid, capsule formation, curli
synthesis, antigen 43 and several transcription factors that can
affect expression of genes outside of these clusters (Figure
\ref{fig:figThree}), such as RcsA, CsgD, SdiA and UvrY
\cite{Beloin2008a}. The change in the local DNA structure and topology
affects these two classes of genes in opposite ways thus contributing
to the transition from a motility to an adherence phenotype
(Supplementary Table \ref{tab:tablex}).

While the role of H-NS in the regulation of these pathways is well
known, the role of Fis is still less well-defined. A large set of
flagellar genes is activated by Fis both under high and low
supercoiling conditions, mostly those found in Cluster 1, while the
expression of csgA is repressed by Fis at low supercoiling as reported
by \cite{Kelly2004, Saldana2009} (Supplementary Table
\ref{tab:tabley}). In addition, Fis seems to mediate the supercoiling
dependence of flhC and rcsA expression, the first being in the
$\Delta$Fis hyp and the second in the $\Delta$Fis rel list. These two
proteins play opposite roles on flagellar synthesis pathway
(Figure~\ref{fig:figThree}).

Finally, Cluster 6 in the Ori macrodomain corresponds to the
chromosomal waa region containing the operons for lipopolysaccaride
(LPS) synthesis necessary for biofilms formation
\cite{Beloin2008a}. The genes found in this cluster respond to changes
in supercoiling in the absence of Fis, overlapping with a cluster of
H-NS binding sites as shown by ChIP-chip analysis, consistent with the
presence of several predicted H-NS sites according to Lang et al
\cite{LBB+07}.
The chromosomal map of the known genes involved in synthesis and
regulation of flagella, chemotaxis and the different stages of biofilm
formation reveals a preferential localization in the Ori and Ter
macrodomains (Figure 3).

To conclude, we believe that this study shows the power of the
integrated analysis of distinct datasets in the context of the role
played by the nucleoid in transcription.  In the future, in order to
more easily integrate datasets from different sources, it will be
necessary to start a common effort towards the construction of larger
comprehensive databases and consortia collecting, sharing and
analyzing data obtained with different high- and low- throughput
techniques.

\clearpage

\section{Methods}

\subsection{Microarray data of Fis, H-NS, and supercoil sensitive genes.}
We used microarray data from Blot~\emph{et~al.}~\cite{blot}, where the
wild type level of gene expression is measured (from cells growing in
rich media) relative to genetically engineered \emph{E.~coli~LZ41} and
\emph{LZ54} strains containing drug-resistant topoisomerase gene
alleles to inhibit DNA gyrase or topoisomerase IV activity
selectively~\cite{zechiedrich} and thereby inducing negative supercoil
relaxation ($-\sigma < 0.033$ on average, as measured on plasmids) or
increase ($-\sigma > 0.08$ on average, as measured on plasmids).
In addition, these strains were crossed with the knockouts $\Delta$Fis
and $\Delta$H-NS in order to determine the coupling of the effect of a
specific nucleoid protein with a specific level of supercoiling.
These experiments define seven sets of nucleoid-perturbation sensitive
genes relative to pairs of conditions compared. The so-called
``intra-strain'' sets include the genes that change significantly
their expression when the negative superhelical density $\sigma$
varies in a fixed genetic background. The ``inter-strain'' sets
include the genes that change significantly their expression at a
fixed average supercoiling background (-$\sigma < 0.033$ or -$\sigma >
0.08$) comparing the transcript profiles of the wild-type and knockout
mutant.

In the text we refer to intra-strain lists by the name of the relative
mutant, WT, $\Delta$Fis and $\Delta$H-NS, and to inter-strain lists by
the names of the two mutants compared, with a suffix indicating the
supercoiling condition, WT-$\Delta$Fis (high or low $-\sigma$),
WT-$\Delta$H-NS (high or low $-\sigma$).  Supplementary Figure
\ref{fig:liste} summarizes both the experimental sets and the gene
lists.

We also considered microarray data on a $\Delta$Fis strain in
different growth phases (early, mid-, late-exponential and stationary,
in rich media) from both Bradley~\emph{et~al.}~\cite{bradley} and
Blot~\emph{et~al.}~\cite{blot} datasets.  The lists of genes
significantly changing their expression with respect to wild type are
referred to by the names of the two mutants compared with a suffix
indicating the growth phase, WT-$\Delta$Fis (early/mid/late/stat
phase) for Bradley data and WT-$\Delta$Fis/H-NS (LS or late stationary
phase/ME or mid-exponential phase/TS or transition phase) for data
from Blot \emph{et al.}.

\subsection{Transcription network.}
The transcription network interactions were compiled from the RegulonDB
6.0 database \cite{GJP+08}, which contains a concise representation
of the information available from the literature about transcriptional
regulation of all genes in \emph{E.~coli}. Interactions inferred
purely from microarrays were filtered out in order to
decrease the contribution of indirect effects. Of 4552 genes in
\emph{E.~coli}, 1524 genes are in the unfiltered network in RegulonDB,
1372 are in the filtered network and of these 166 and 140
respectively are transcription factors, out of the 286 genes
functionally annotated as transcription factors in \emph{E.~coli} by
Gene Ontology \cite{ecocyc, go}. 

\subsection{ChIP-chip binding profiles and protein occupancy data.}
Binding sites from ChIP-chip data of H-NS and Fis proteins along the
\emph{E.~coli} genome \emph{in vivo} were obtained from
Grainger~\emph{et~al.}~\cite{grainger}. These sets are referred to as
FISbind(Grainger), HNSbind(Grainger) in the text. As a comparison, we
also considered the list of Fis binding sites obtained by high-density
ChIP-chip by Cho~\emph{et~al.}~\cite{cho2}, identifying 894
Fis-associated binding regions (compared to the 224 regions found by
Grainger~\emph{et~al.}), referred to as FISbind(Cho), and the list of
H-NS binding data from Oshima \emph{et al.}~\cite{Oshima2006},
HNSbind(Oshima).
We also considered data from RegulonDB for Fis and H-NS binding sites,
referred to as FISbind(RegulonDB) and HNSbind(RegulonDB) in the text.
In order to quickly compare the overlap between RegulonDB target genes
and Fis and H-NS ChIP-chip data sets we have carried out a
hypergeometric test with results reported in Supplementary Table
\ref{tab:bindhyp}. Both ChIP-chip data sets refer to cells grown in
minimal media.

Finally, the data from Vora \emph{et~al.} (ref.~\cite{vora}) was
considered.  In this study the amount of total protein-DNA
interactions at a specific locus \emph{in vivo} was measured (from
cells grown in rich media) by a modified large-scale ChIP assay
measuring generic protein occupancy along the genome (termed \emph{in
  vivo} protein occupancy display, IPOD). Specifically, we examined
the ($>1$ Kb) protein occupancy domains (EPODs), divided by the
authors into two populations by their median expression level (121
domains in the highly expressed class, heEPODs, and 151 in the
transcriptionally silent class, tsEPODs).

\subsection{Statistical analysis of spatial clusters.}
We developed a statistical method for identifying clusters of genes in
the lists along the genomic coordinate. This method considers the density
of genes at different scales on the genome, and compares empirical
data with results from random null models.
In order to avoid spurious effects of binning, for each gene list a
density histogram was made by using a sliding window with a given
bin-size $b_s$ as exemplified in Figure~\ref{fig:histclust}.  The
resulting plot of the averaged density of genes for every point of the
circular chromosome was considered at different observation scales of
the genome, i.e. at different bin sizes of length $b_s \in \{L/2, L/4,
\dots, L/2^n\}$ where $L$ is the length of the chromosome. We chose
$n=10$, as $b_s < L/1024$ is the scale of the typical gene length.

Density peaks with a significantly high number of genes (see also
Figure~\ref{fig:histclust}) were identified by comparing empirical
data with 10000 realizations of a null model. For every bin size, the
null model considers the density histogram from a random list of the
same length of the empirical one.
The number of genes for every bin in the empirical histogram was
compared to the distribution of global maxima of the null model,
obtaining a $P$-value for the value of the empirical histogram for
each bin. This procedure enables to extract a list of statistically
significant ($P < 0.05$) bin positions. For each bin-size (or
observation scale), clusters were defined as connected intervals
containing a significantly high proportion of the genes in the list.
To each cluster, we assigned the lowest $P$-value among the merged
bins.

\subsection{Macrodomains and chromosomal sectors.}
The location of the chromosomal macrodomains were extracted
from~\cite{VPR+04,boccard1,EB06}, and considered together with the
chromosomal sectors of ref.~\cite{Mathelier2010} where codon bias
indices positively correlate with gene expression. The exact
coordinates used here are presented in Supplementary Table
\ref{tab:macrod}

\subsection{Periodicity analysis.}

Periodic signals in the position of genes of a given list were derived
from both the density (histogram of the start position of each gene)
and the histogram of the shortest distances along the genome between
any gene pair.
We computed the discrete Fast Fourier transform of this function. For
every frequency $\nu$, the spectra of the Fourier transform is
proportional to the strength of the periodic signal of period $\lambda
= 1/\nu$, while the complex phase $\theta$ is proportional to a shift
of the periodic distribution $\tau = \theta \lambda / 2 \pi$ with
respect to the cosinusoidal periodic distribution of that period.
Note that the resolution of this analysis is limited to a few bin
sizes, so that since the distance distribution between gene pairs
contains more data points, it allows to probe more effectively smaller
length scales.  Being $L$ the number of bases in the chromosome (the
maximal distances is $L/2$) we used a bin-size of $L/256$ bases for
the position histogram and a bin-size of $L/2048$ bases for the
distance histogram.

Peaks exhibited by empirical data were scored with the same null model
used for the cluster analysis, i.e. randomized lists of genes
conserving the length of the empirical list. Given an empirical list,
we generated 500 random lists of the same length. The lower sampling
compared to the 10000 random lists generated for the clustering is due
to the fact that the pair distance distribution requires the storage
and elaboration of $n^2$ data points instead of $n$, which causes a
consequent increase of computing time. The height of the spectral
peaks for every periodicity in the empirical distributions was
compared to the distribution of global maxima of the null model
(regardless of their position in the spectrum) as exemplified in
Supplementary Figure~\ref{fig:figperi}, a periodicity with a $P$-value
$< 0.05$ was considered to be statistically significant.

\subsection{Functional annotations.}
Functional annotations were downloaded from the MultiFun web site
\url{http://genprotec.mbl.edu/}~\cite{Serres2000}.  Gene sets
belonging to clusters and effective networks were probed for
enrichment of functional annotations by hypergeometric testing.
$P$-values lower than $10^{-3}$ were considered significant.

\section*{Acknowledgments}
The authors acknowledge support from the Human Frontier Science
Program Organization (Grant RGY0069/2009-C). We are also very grateful
to A.~Travers, G.~Muskhelishvili, and V.G.~Benza for critical reading
of this manuscript and to O.~Espeli, A.~Carbone, I.~Junier and
A.~Mathelier for useful discussions.

\bibliography{mdclusbibf} 
\bibliographystyle{rsc} 
\clearpage

 \fancyhead[L]{}
 \fancyhead[C]{}
 \fancyhead[R]{}
\fancyfoot[LE]{\footnotesize{\sffamily{\thepage}}}
\fancyfoot[RO]{\footnotesize{\sffamily{\thepage}}}
 \fancyfoot[LO,RE]{}
 \fancyfoot[CO]{}
 \fancyfoot[CE]{}

\setcounter{page}{1}
{ \center
\section*{\Large Supplementary methods for Scolari \emph{et al}} }

\renewcommand{\thesection}{Supplementary Methods \arabic{section}}
\setcounter{figure}{0} 
\setcounter{table}{0} 
\setcounter{section}{0}
\renewcommand{\tablename}{Supplementary Methods Table}
\renewcommand{\figurename}{Supplementary Methods Figure}
\renewcommand{\thefigure}{SM\arabic{figure}}
\renewcommand{\thetable}{SM\arabic{table}}


\begin{figure}[H]
  \centering
  \includegraphics[width=0.4\textwidth]{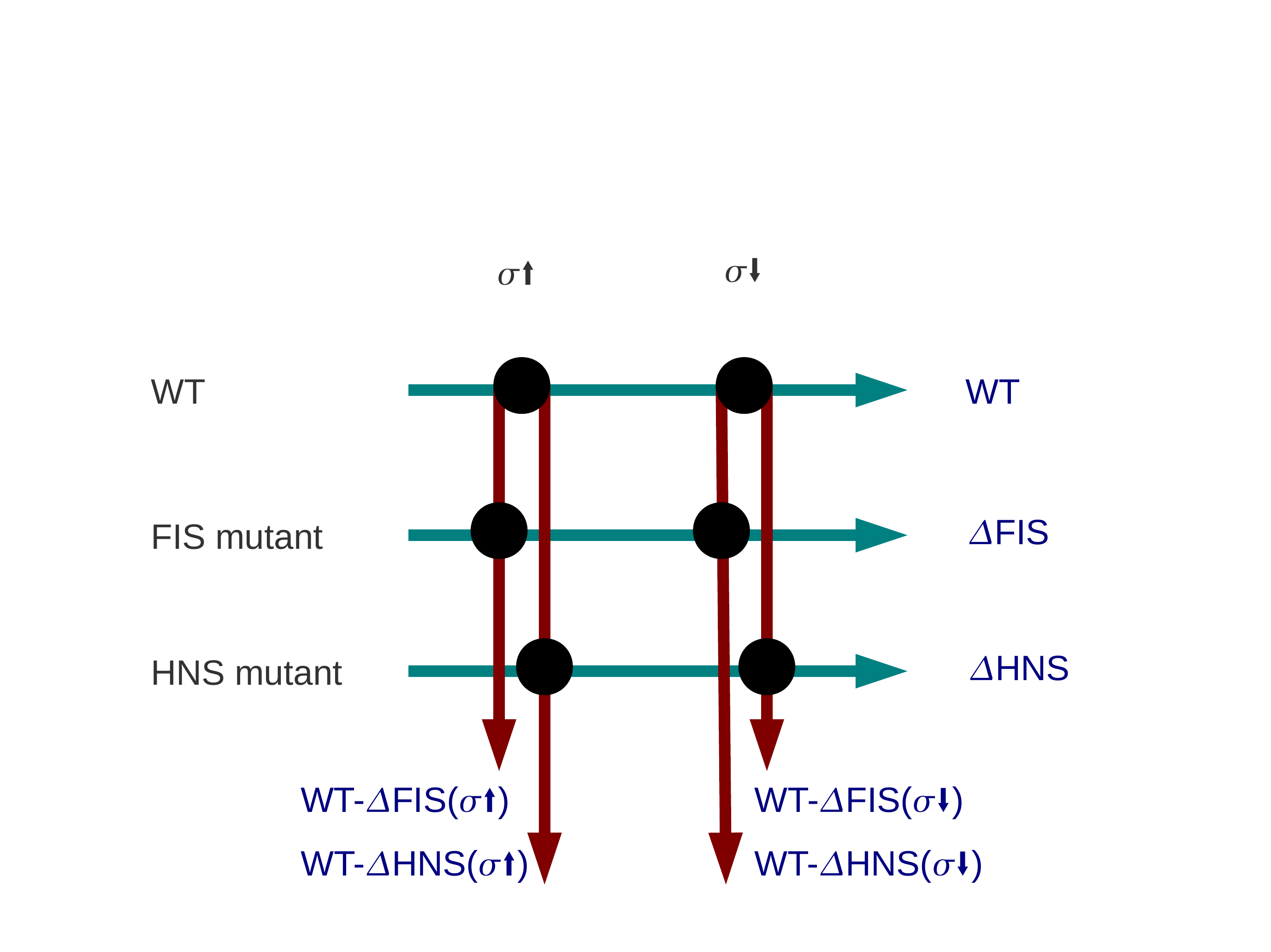}
  \caption{Summary of microarray data of Fis, H-NS, and supercoil
    sensitive genes.}
\label{fig:liste}
\end{figure}


\begin{table}[H]
\begin{minipage}{.5cm}
{\bf (a)}
\end{minipage}
\begin{minipage}{10cm}
\begin{tabular}{r|ccc}
  & RegulonDB & Cho  \emph{et al}          &    Grainger \emph{et al}    \\
  \hline
  RegulonDB  &   $200$   & $58 \  (P = 10^{-4})$ & $11 \  (P = 0.26)$    \\
  Cho        &           &  $894$         & $71 \  (P = 10^{-9})$ \\
  Grainger   &           &                &     $220$      \\
\end{tabular}
\end{minipage}
\begin{minipage}{.5cm}
{\bf (b)}
\end{minipage}
\begin{minipage}{7cm}
\begin{tabular}{r|cc}
           & RegulonDB & Grainger  \emph{et al} \\
\hline
RegulonDB  &   $71$      &  $1 \  (P = 0.44)$ \\
Grainger   &           &   $96$      \\
\end{tabular}
\end{minipage}
\caption{The table compares reported Fis {\bf (a) } and H-NS {\bf (b) }
  binding    sites   from different data sources. The table reports
  the number of genes in the overlap between the lists, and the
  $P$-value in parentheses.  } 
\label{tab:bindhyp}
\end{table}


\begin{table}[H]
\centering
  \begin{minipage}{1cm}
    {\bf (a) }
  \end{minipage}
  \begin{tabular}{lcccccccc}
    \toprule
    \bf{Macro-}&\multicolumn{2}{c}{{\bf Ori}} &
    \multicolumn{2}{c}{{\bf Right}} & \multicolumn{2}{c}{{\bf Ter}} &
    \multicolumn{2}{c}{{\bf Left}}\\
    \cmidrule(lr){2-3} \cmidrule(lr){4-5} \cmidrule(lr){6-7}
    \cmidrule(lr){8-9}  
    \bf{domain:}& start & end & start & end & start & end & start & end \\
    \midrule 
    Position (\emph{Mb}) &
    3.76 & 0.05 &
    0.60 & 1.21 &
    1.21 & 2.18 &
    2.18 & 2.88 \\
    \bottomrule
  \end{tabular}
  \vspace{.5cm}\\
  \begin{minipage}{0.5cm}
    { \bf (b) }
  \end{minipage}
  \begin{tabular}{lcccccccccccccc}
    \toprule
    \bf{Chromosomal}&\multicolumn{2}{c}{{\bf A}} &
    \multicolumn{2}{c}{{\bf G}} & \multicolumn{2}{c}{{\bf F}} &
    \multicolumn{2}{c}{{\bf E}} & \multicolumn{2}{c}{{\bf D}} &
    \multicolumn{2}{c}{{\bf C}} & \multicolumn{2}{c}{{\bf B}}\\
    \cmidrule(lr){2-3} \cmidrule(lr){4-5} \cmidrule(lr){6-7}
    \cmidrule(lr){8-9}
    \cmidrule(lr){10-11} \cmidrule(lr){12-13} \cmidrule(lr){14-15}
    \bf{sector:}& start & end & start & end & start & end & start & end
    & start & end & start & end & start & end\\
    \midrule 
    Position (\emph{Mb}) &
    3.59 & 0.59 &
    0.68 & 1.33 &
    1.40 & 1.49 &
    1.54 & 1.97 &
    2.03 & 2.16 &
    2.27 & 2.86 &
    2.97 & 3.57 \\
    \bottomrule
\end{tabular}
\caption{{\bf (a)} Start and end positions in $Mb$ of the
  macrodomains defined by Boccard and coworkers
  (ref.~\cite{VPR+04,EMB08}),  and {\bf (b)} chromosomal sectors
  defined by Mathelier and Carbone (ref.~\cite{Mathelier2010})} 
\label{tab:macrod}
\end{table}

\clearpage

\setcounter{page}{1}

{\center
\section*{\Large Supplementary results for Scolari \emph{et al}}}

\renewcommand{\thesection}{Supplementary Section \arabic{section}}
\setcounter{figure}{0} 
\setcounter{table}{0} 
\setcounter{section}{0}
\renewcommand{\tablename}{Supplementary Table}
\renewcommand{\figurename}{Supplementary Figure}
\renewcommand{\thefigure}{S\arabic{figure}}
\renewcommand{\thetable}{S\arabic{table}}

\subsubsection*{Clusters from nucleoid perturbation / transcriptomics
  experiments}\qquad


\begin{figure}[h]
  \centering
  \includegraphics{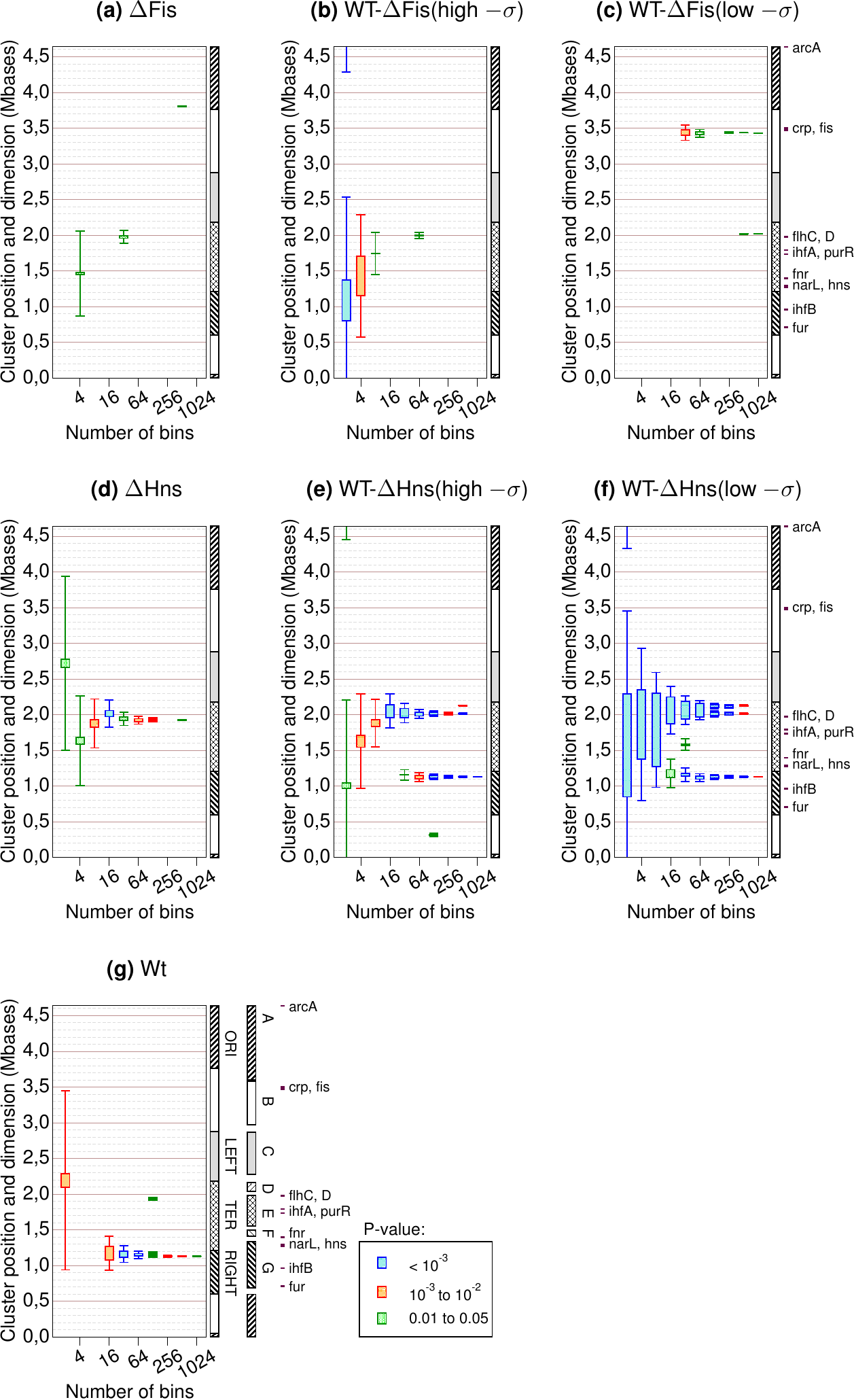}
  \caption{Cluster diagrams for the transcription microarray nucleoid
    perturbation data from ref.~\cite{blot,MGH+08}. }
\label{fig:cluexpr}
\end{figure}

\begin{figure}[H]
  \centering
  \includegraphics{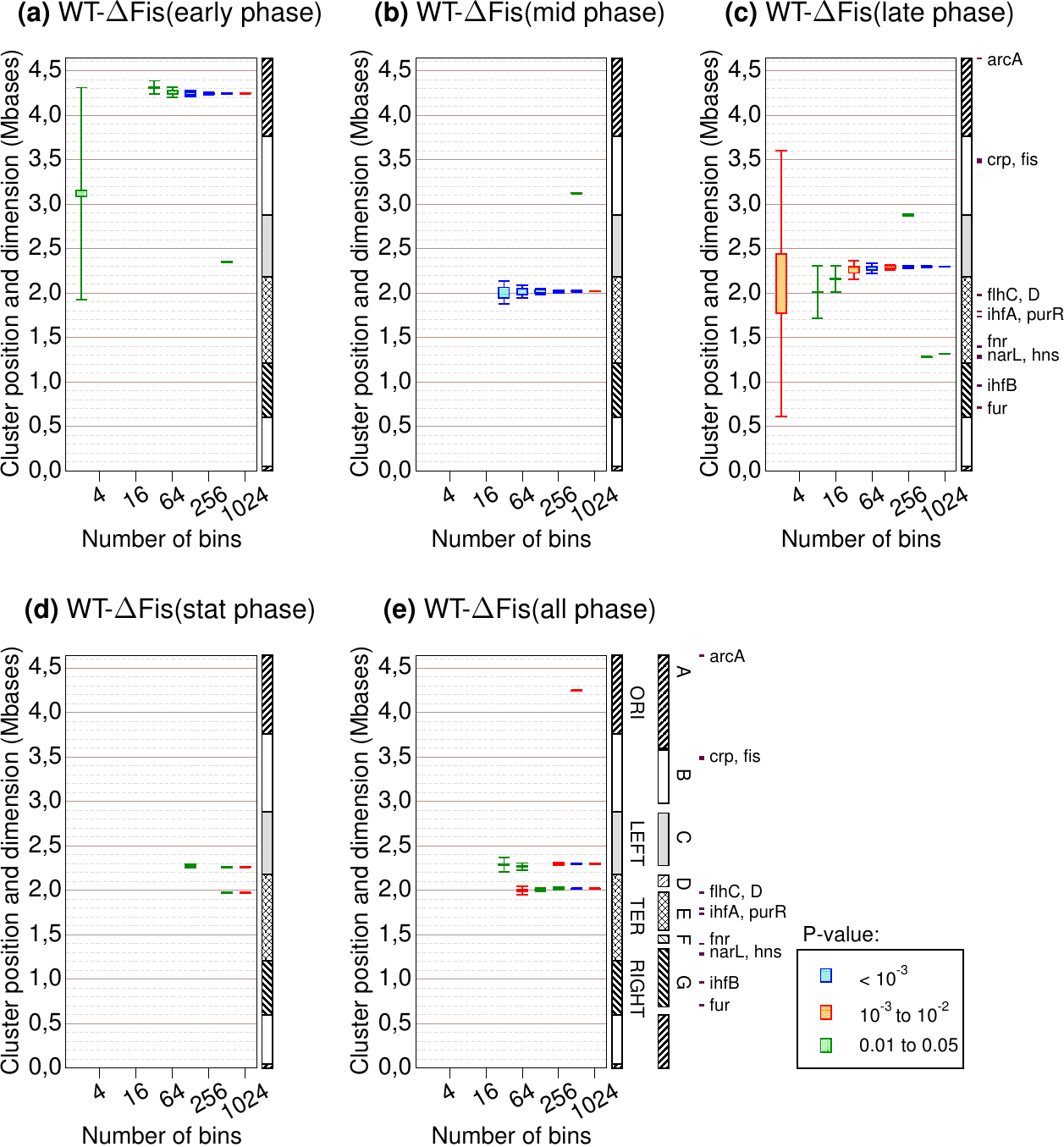}
  \caption{Cluster diagrams for the transcription microarray Fis
    deletion data from ref. \cite{bradley}. Different panels refer to
    different growth phases (early, mid-, late-exponential and
    stationary, in rich media), while the last panel refers to the
    union of all the growth phases. }
\label{fig:blubrad}
\end{figure}

\clearpage
\newpage

\subsubsection*{Clusters from protein binding data:}\qquad


\begin{figure}[h]
  \centering
  \includegraphics{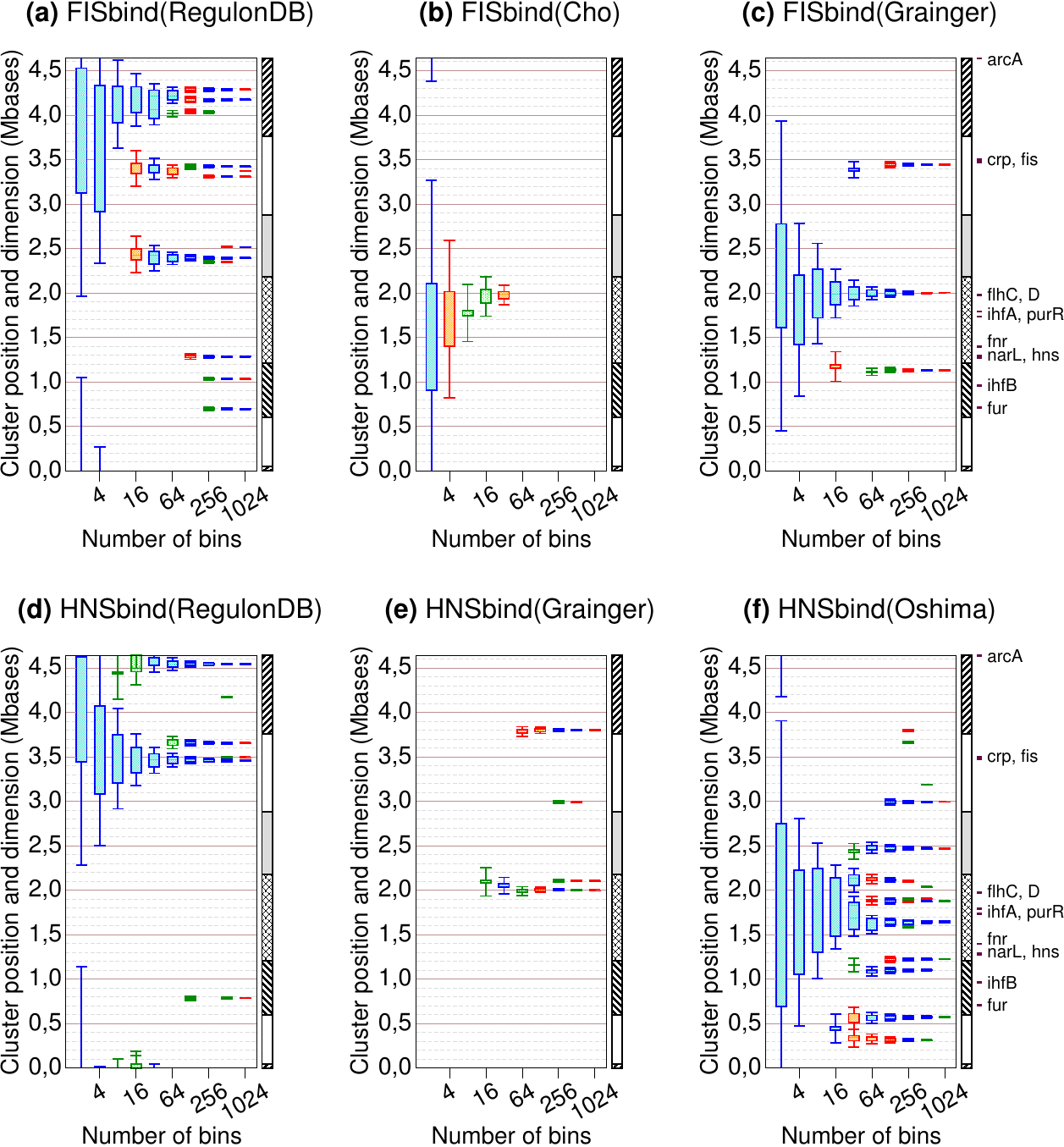}
  \caption{Cluster diagrams of Fis and H-NS binding from the
    RegulonDB~\cite{GJP+08} database and Grainger, Cho and Oshima ChIP-chip
    experiments~\cite{grainger,cho2,Oshima2006}. The binding sites of the
    Grainger data-set for the Fis and H-NS experiments show the same
    clusters as transcriptomics data on genes responding to
    supercoiling changes after H-NS deletion in microarray data from
    Marr \emph{et al.}~\cite{blot,MGH+08}. A hypergeometric test was
    performed to test the overlaps of the Grainger and Cho ChIP-chip
    targets with the RegulonDB database (Table \ref{tab:bindhyp})}.
\label{fig:clubind}
\end{figure}
\clearpage
\subsubsection*{Summary of clusters from transcriptomics and protein
  binding data:}\qquad

\begin{table}[h!]
\centering
\begin{tabular}{r||c|c|c||c||c||c||c}
\multicolumn{8}{c}{From transcriptomics:} \\
  list & {\bf clust1} & {\bf clust2} & {\bf clust3} & {\bf clust4} &
  {\bf clust5} & {\bf clust6} & {\bf clust7} \\
\hline
\hline
  WT    & $0.03$ &  &   &  $< 0.001$  &  &  &    \\
\hline
  $\Delta$Fis   & $0.05$ & & & & & $0.04$ &  \\ 
\hline
  $\Delta$H-NS   & $0.01$ & & & & &      & \\
\hline
  WT-$\Delta$Fis(low) & $0.04$ & $0.04$ & & & $0.01$ &  \\
\hline
  WT-$\Delta$Fis(high) & $0.05$ & $0.06$ & & & &  \\
\hline
  WT-$\Delta$H-NS(low) & $< 0.001$& $< 0.001$& $< 0.001$& $< 0.001$ & 
  & & \\ 
\hline
  WT-$\Delta$H-NS(high) & $< 0.001$& $< 0.001$& $ 0.01$& $< 0.001$ & 
  & & \\ 
\hline
\multicolumn{8}{c}{From protein binding data:} \\
  list & {\bf clust1} & {\bf clust2} & {\bf clust3} & {\bf clust4} &
  {\bf clust5} & {\bf clust6} & {\bf clust7} \\
\hline
  GraingerFis & $< 0.001$ & $< 0.001$ & & $< 0.001$ & $< 0.001$ & & \\
\hline
  GraingerHns & $< 0.001$ & $< 0.001$ & $0.01$ & & & $< 0.001$ &
  $0.01$\\
\hline
\end{tabular}
\caption{Summary of the most significant clusters found at all scales and their $P$-values. The coordinates
  (in bp) along the \emph{E.~Coli} genome are:
  {\bf Cluster 1} 1929600-2195230
  {\bf Cluster 2} 1993030-2037780
  {\bf Cluster 3} 2096110-2141990
  {\bf Cluster 4} 1094210-1163310
  {\bf Cluster 5} 3428200-3447460
  {\bf Cluster 6} 3782180-3815590
  {\bf Cluster 7} 2981340-2996630. Note: {\bf Cluster 2} and {\bf Cluster 3} are
  included in {\bf Cluster 1}.
}
\label{tab:space}
\end{table}

\clearpage
\subsubsection*{Clusters of genes inside and outside the known
  transcription regulatory network:}\qquad

\begin{figure}[H]
  \centering
  \includegraphics{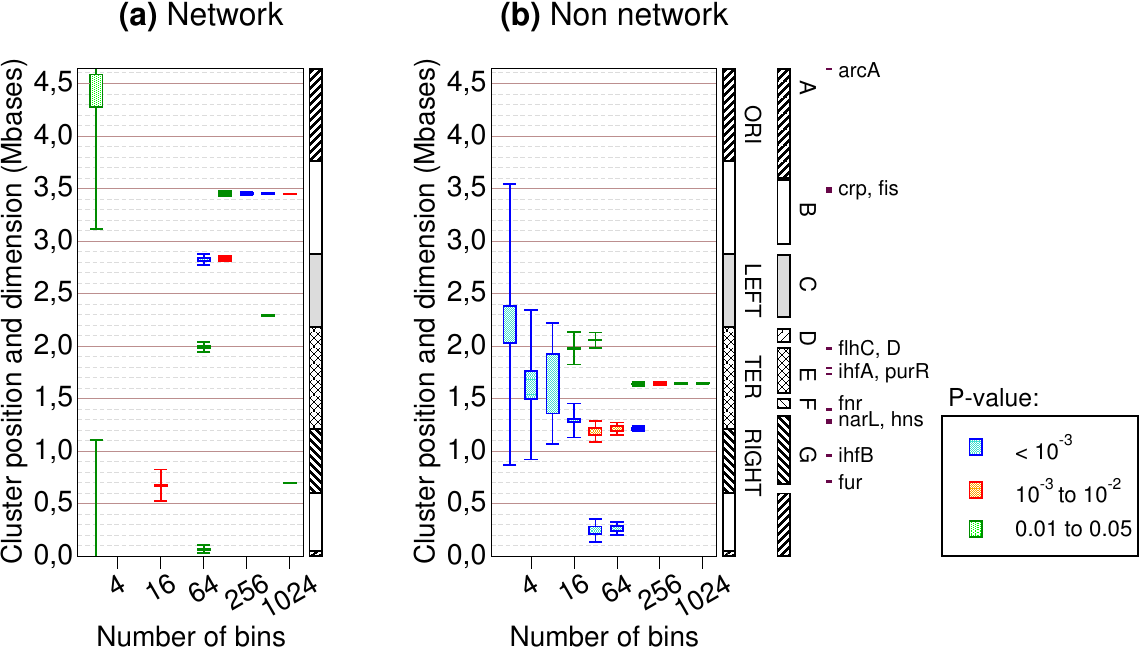}
  \caption{Cluster diagram of the genes in the RegulonDB network and
    outside RegulonDB network (see ref.~\cite{Sonnenschein2009}).}
\label{fig:netdiagram}
\end{figure}

\clearpage

\subsubsection*{Histograms of EPODs and comparison with clusters:}\qquad

\begin{figure}[H]
  \centering
  \includegraphics{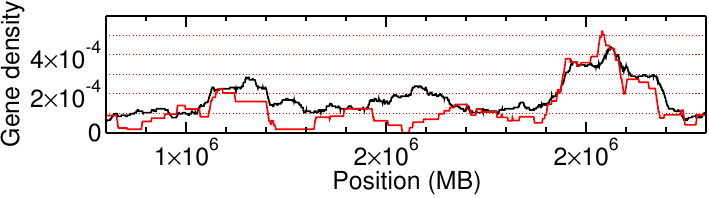}
  \caption{Linear density of heEPODs along the genome (red line)
    compared to the density of nucleoid-perturbation sensitive genes
    WT-$\Delta$H-NS(low $-\sigma$) (black line, bin size $L/32$). The
    x-axis spans the Ter macrodomain. Note the highly correlated
    regions at the border of the Ter (at $1\cdot10^6$ and $2\cdot10^6$
    bases) macrodomain in accordance to the results of
    Figure \ref{fig:epodcorr}.  }
\label{fig:epodsuper}
\end{figure}

\begin{figure}[H]
  \centering
  \includegraphics{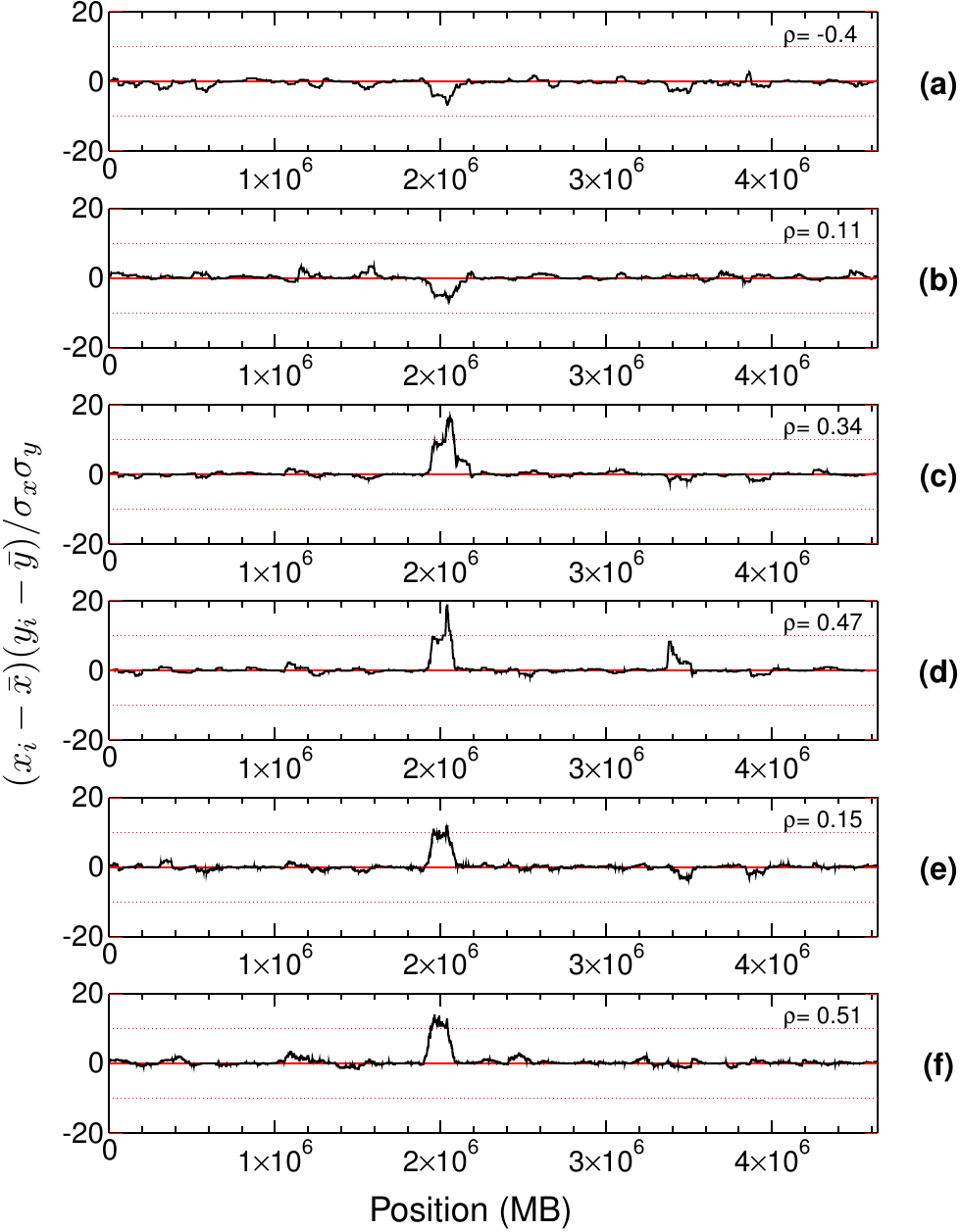}
\label{fig:A3}
\caption{The black lines are the local contributions to the Pearson
  correlation coefficient along the genome coordinate (x-axis) between
  the normalized linear densities of he- and tsEPODS
  (ref.~\cite{vora}) and the densities of nucleoid-perturbation
  sensitive genes. The densities where calculated using a sliding
  window of size $L/32$.  {\bf(a)}~Correlation between tsEPOD and
  heEPOD density.  {\bf(b)}~Correlation between tsEPOD density and
  WT-$\Delta$H-NS(low $-\sigma$).  {\bf(c)}~Correlation between heEPOD
  and WT-$\Delta$H-NS(low $-\sigma$).  {\bf(d)}~Correlation between
  heEPOD density and FISbinding(Grainger).  {\bf(e)}~Correlation
  between heEPOD and FISbinding(Cho).  {\bf(f)}~Correlation between
  FISbinding(Grainger) and FISbinding(Cho).  }
\label{fig:epodcorr}
\end{figure}

\clearpage

\begin{table}[H]
\centering
\vspace{5mm}
\begin{tabular}{cccc}
\toprule
\multicolumn{4}{c}%
{\textbf{Number of genes of heEPOD in common with:}} \\
\cmidrule(l){1-4}
List          &   Experimental value & Mean value & $P$-value \\
\midrule
WT            &   19  &   16.58     &   0.300 \\
$\Delta$FIS   &   11  &   12.98     &   0.226 \\
$\Delta$HNS   &   15  &   9.06      &   0.034 \\
WT-$\Delta$FIS($\sigma\downarrow$)   &  22 & 13.29 & 0.004 \\
WT-$\Delta$HNS($\sigma\downarrow$)   &  31 & 15.64 & $<$ 0.002 \\
WT-$\Delta$FIS($\sigma\uparrow$)     &  18 & 8.69  & $<$ 0.002 \\
WT-$\Delta$HNS($\sigma\uparrow$)     &  30 & 11.52 & $<$ 0.002\\
\bottomrule
\end{tabular}
\caption{Summary of the number of genes within heEPODs  (strictly
  included) in common with the  genes significantly 
  responding to Fis and H-NS deletion, and changes in
  supercoiling~\cite{blot}.}   
\label{tab:overlap_he_zero}
\end{table}

\begin{table}[H]
\centering
\vspace{5mm}
\begin{tabular}{cccc}
\toprule
\multicolumn{4}{c}%
{\textbf{Number of genes of tsEPOD in common with:}} \\
\cmidrule(l){1-4}
List          &   Experimental value & Mean value & $P$-value \\
\midrule
WT            &   5  &   6.16     &   0.224 \\
$\Delta$FIS   &   7  &   5.43     &   0.298 \\
$\Delta$HNS   &   6  &   5.65     &   0.496 \\
WT-$\Delta$FIS($\sigma\downarrow$)   &  5  & 4.68  & 0.496 \\
WT-$\Delta$HNS($\sigma\downarrow$)   &  13 & 5.70  & 0.002 \\
WT-$\Delta$FIS($\sigma\uparrow$)     &  4  & 3.20  & 0.402 \\
WT-$\Delta$HNS($\sigma\uparrow$)     &  13 & 4.35  & $<$ 0.002\\
\bottomrule
\end{tabular}
\caption{Summary of the number of genes within tsEPODs  (strictly
  included) in common with the  genes significantly 
  responding to Fis and H-NS deletion, and changes in supercoiling~\cite{blot}.}
\label{tab:overlap_ts_zero}
\end{table}

\clearpage
\subsubsection*{Clusters of genes controlled by FlhC:}\qquad

\begin{figure}[H]
\centering
  \includegraphics[height=8cm]{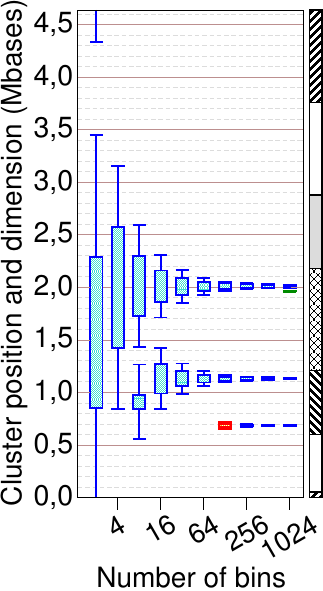}
  \caption{Cluster diagram of the genes controlled by the FlhC
    transcription factor (data from RegulonDB). FlhC is a
    transcriptional activator that controls the operons related to
    assembling of the flagella. The main regions controlled by FlhC
    overlaps with the clusters at the border of the Ter macrodomain
    identified in both transcriptomics and binding sites lists. A
    second cluster is present in correspondence with the border
    between the Right macrodomain and the Right non structured zone.
  }
\label{fig:flaclust}
\end{figure}

\clearpage

\subsubsection*{Periodicity analysis:}\qquad

\begin{figure}[H]
  \centering
  \includegraphics[width=\textwidth]{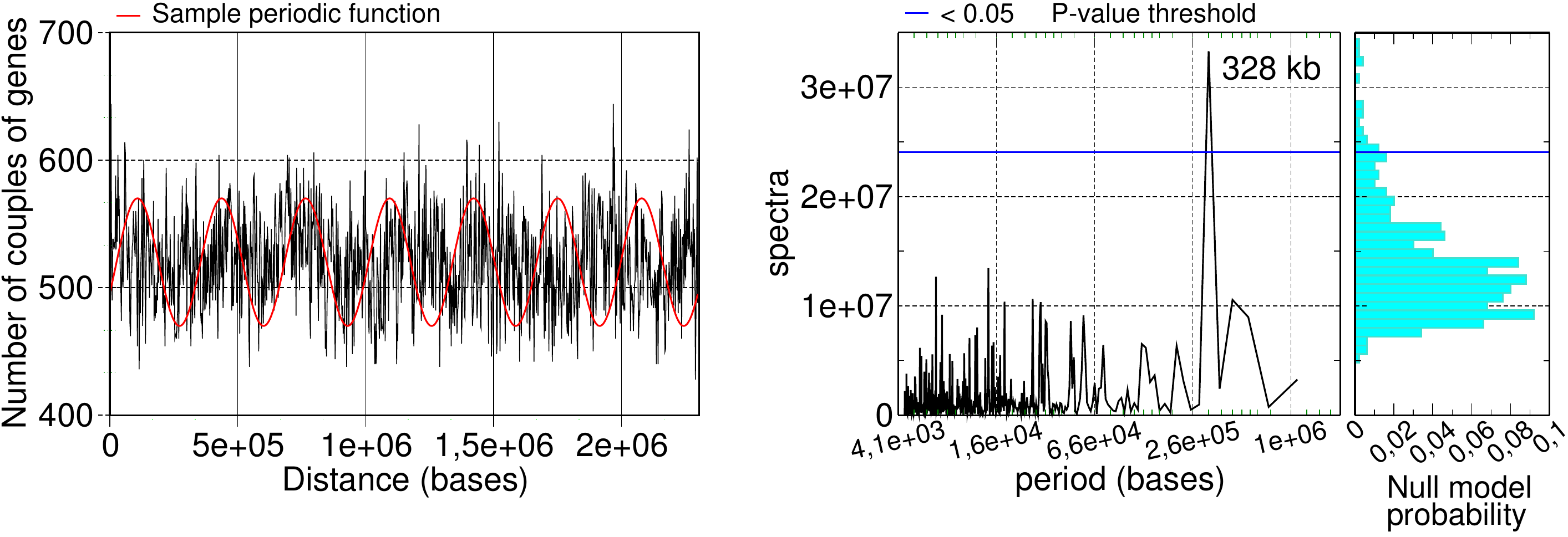}
  \caption{Procedure followed to detect periodicities in the gene
    lists. Left panel: distribution of the distance of genes in an
    experimental list (sliding window of bin-size $L/2048$), compared
    to a periodic function. Right panel: discrete Fourier transform of
    the distance distribution. The peaks correspond to contributions
    of a periodic function of a given period, reported on the
    x-axis. The figure shows the spectra of the intra-strain WT list
    (see Methods), where the peak indicates a signal for a periodicity
    of 328Kb. The comparison of this signal with the distribution of
    the maximum of the spectra found in randomized lists gives a
    significance score for this periodicity. The same procedure can be
    applied also to the sliding-window density histogram at a given
    bin size.  }
\label{fig:figperi}
\end{figure}

\begin{figure}[H]
  \centering
  \includegraphics[width=\textwidth]{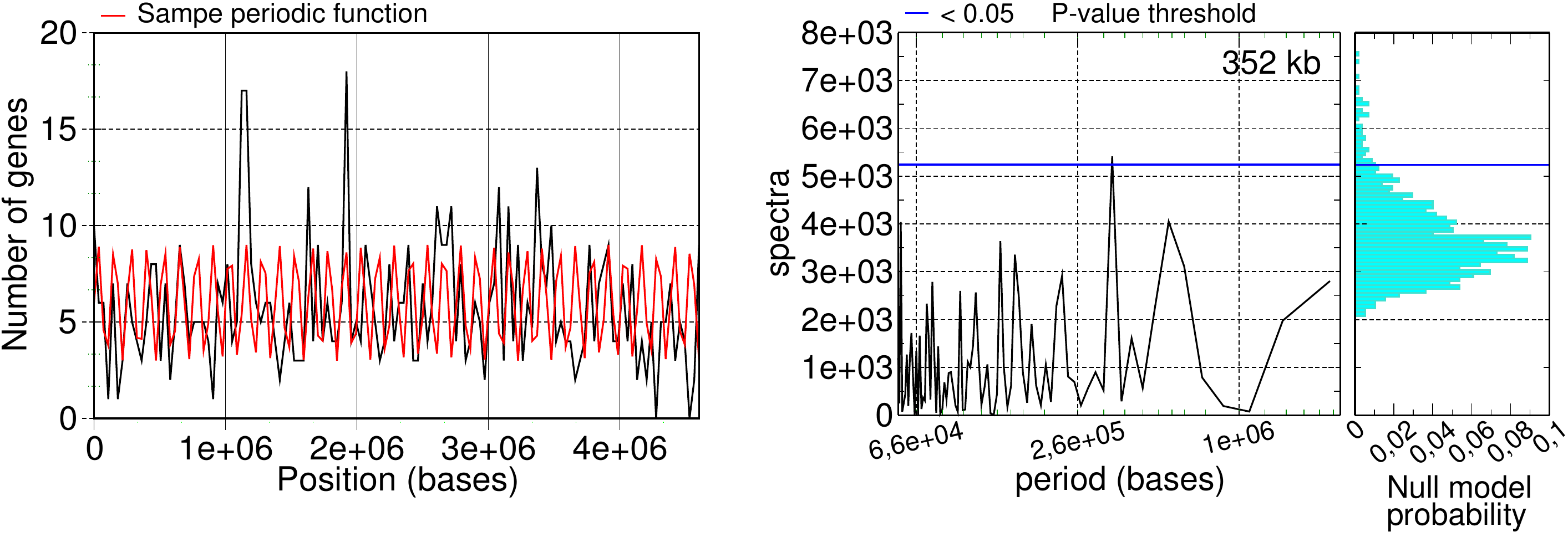}
  \caption{Periodicity analysis for the case of a density
    histogram. The analysis (and the example dataset from the
    intra-strain WT list) coincides with that presented in
    Figure~\ref{fig:figperi}, except that the left panel is a
    sliding-window histogram (bin-size $L/256$) of the genes in the
    list, rather than a distance distribution.}
\label{fig:figperi_pos}
\end{figure}



\begin{table}[H]
\centering
  \begin{tabular}{r||c|c|c|c}
  list & $360 \pm 36$ Kb & $624 \pm 36$Kb & $101 \pm 36$Kb & $20 \pm 36$Kb \\
\hline
\hline
  WT    & AB &   &   &   \\
\hline
  $\Delta$Fis   &   &   & AB &   \\
\hline
  $\Delta$H-NS   &   &   &   & B \\
\hline
  WT-$\Delta$Fis(low)  &   &   &   &   \\
\hline
  WT-$\Delta$Fis(high)  &   &   &   &   \\
\hline
  WT-$\Delta$H-NS(low)  & AB & AB & B &   \\
\hline
  WT-$\Delta$H-NS(high)  & AB & AB & B & B \\
  \end{tabular}
  \caption{Table summarizing the significant periodicities found (P $<
    0.05$). In the table the letter A indicates a significant
    periodicity found in the density histogram while B indicates a
    periodicity found in the  distance pair distribution. Genes
    sensitive to supercoiling variation in the intra-strain WT list
    show a compatible periodicity of $360 \pm 36$Kb, this periodicity
    is found also in 
    the WT-$\Delta$H-NS lists at all supercoiling conditions. Upon Fis
    deletion, supercoiling sensitive genes lose the $360 \pm 
    36$ Kb periodicity, but show a new periodicity at $101 \pm 36$Kb,
    again found also in the WT-$\Delta$H-NS lists in all supercoiling
    conditions. Finally, in H-NS deletion mutants, supercoiling
    sensitive genes lose the $360 \pm 36$ Kb periodicity but show a
    periodicity at $20 \pm 36$Kb also found in the inter-strain
    WT-$\Delta$H-NS data in high negative supercoiling conditions. The
    compatibility condition of $36$Kb was selected as twice the bin-size
    of the density distribution histogram.}
  \label{tab:period}
\end{table}

\clearpage

\subsubsection*{Clusters and the flagellar/biofilm synthesis pathway:}\qquad

\renewcommand{\thesection}{Supplementary Section \arabic{section}}


\renewcommand{\figurename}{Supplementary Table}
\setcounter{figure}{\arabic{table}}

\begin{figure}[H]
  \centering
  \includegraphics[height=17.5cm]{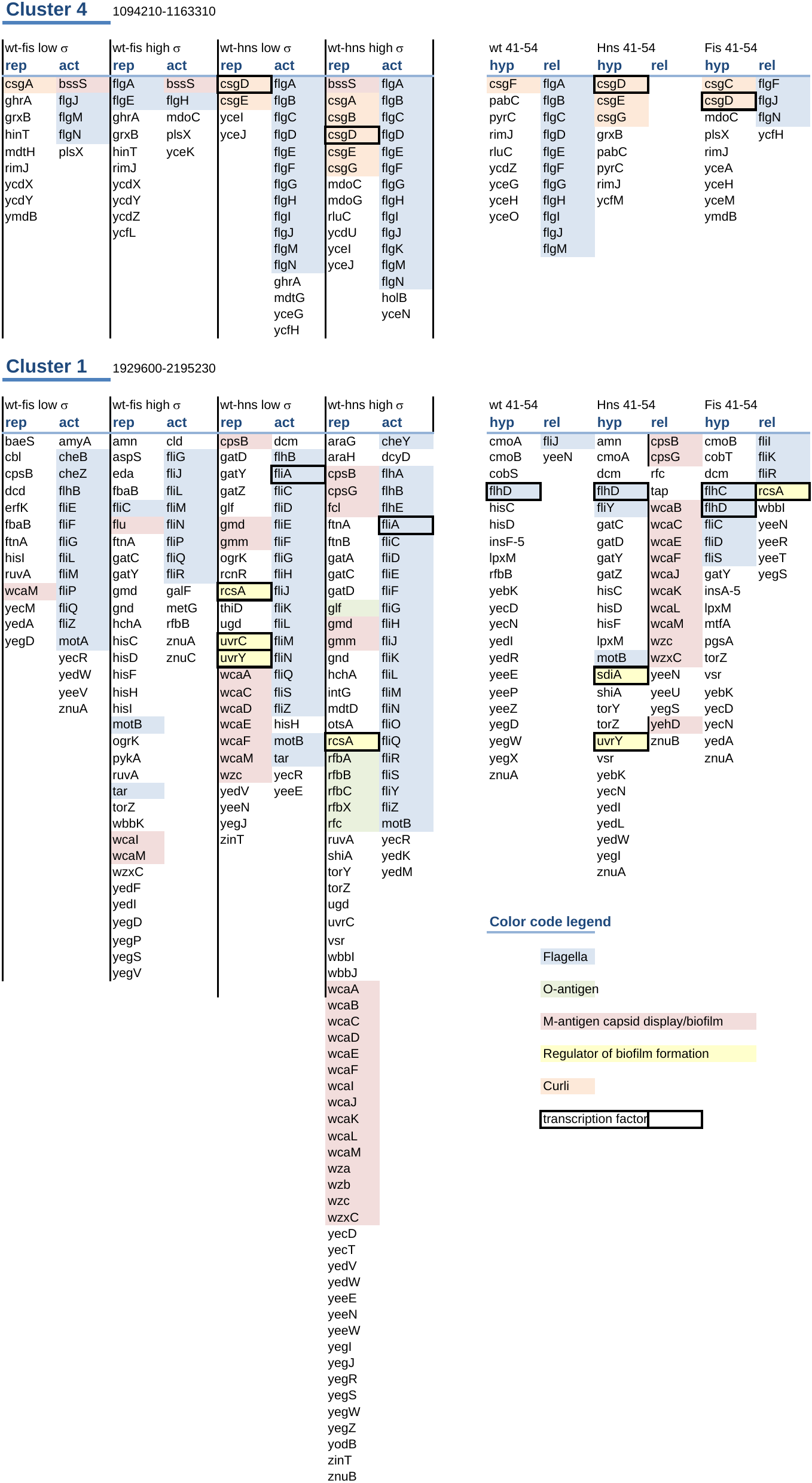}
\caption{Intersection between genes found in the data sets
from the transcription microarray experiments of Blot et
al~\cite{blot} and clusters of genes identified in this analysis. The 
colors indicate the gene ontology class. Genes from intra-strain
experiment have been divided into rel and hyp columns corresponding to
gene transcripts whose expression is associated with relaxation (rel)
or high negative supercoiling (hyp). The labels act and rep indicate
activation and repression in inter-strain profiles.}
\label{tab:tablex}
\end{figure}

\begin{figure}[H]
  \centering
  \includegraphics[height=18cm]{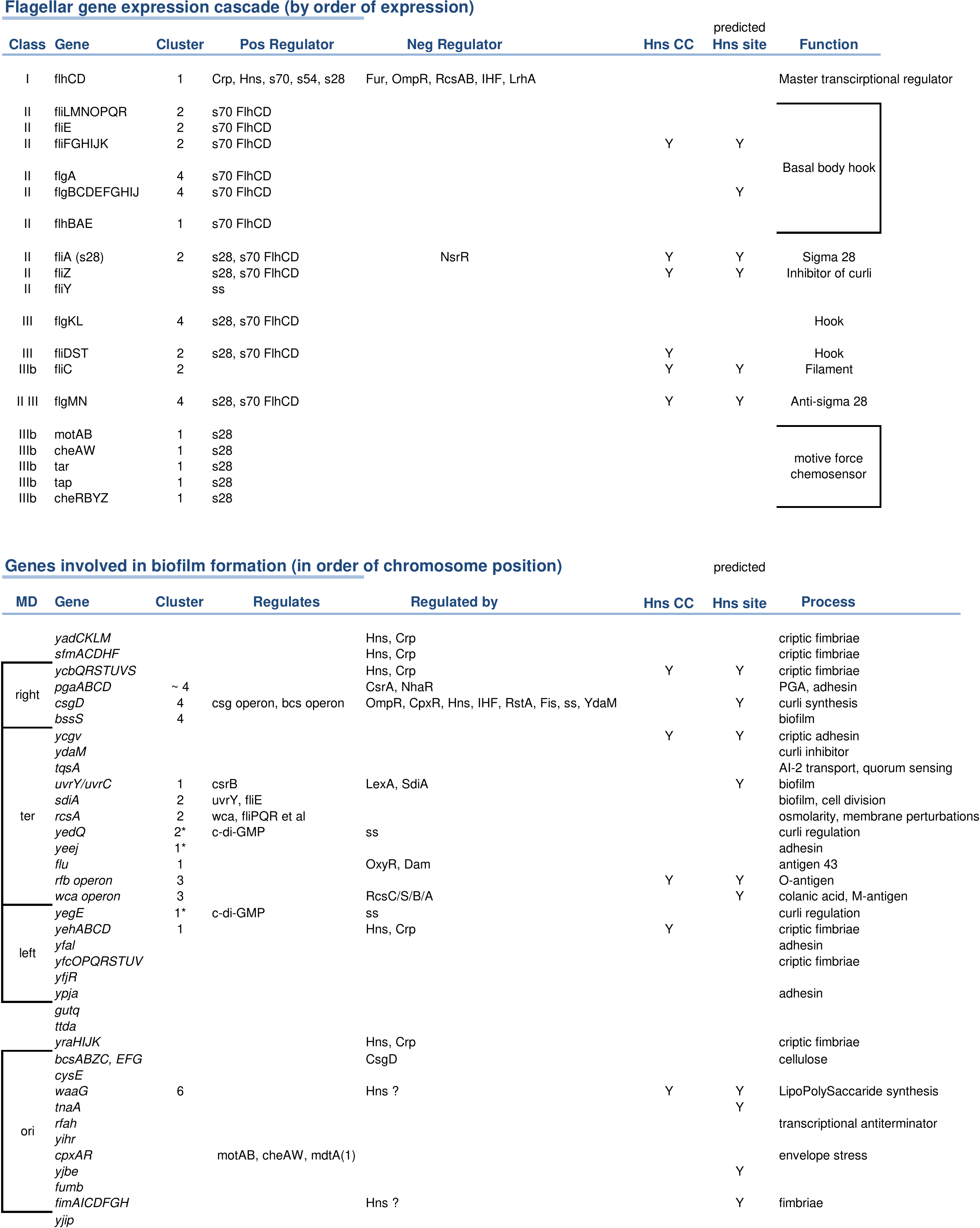}
\caption{Genes involved in flagellar expression and in biofilm
formation, data from refs
\cite{Beloin2008a,Hengge2009b,Ko2000,Krin2010,Ogasawara2010,Saldana2009,Wang2004,Korea2010,Prigent-Combaret2001}. Flagellar
and chemotaxis genes are ordered according to the sequence of
expression. Biofilm genes are ordered according to their position on
the chromosome, the first column (MD) show the macrodomain in which
the gene is located on the chromosome. The known factors regulating
gene expression are indicated, as well as the targets of the
transcription factors in the list, the abbreviation ss stands for
sigma s. The presence of H-NS binding sites in the promoter region is 
shown in different columns whether it was determined by
ChIP-chip~\cite{grainger} (Hns CC) or by prediction from bioinformatic
sequence analysis~\cite{LBB+07} (Hns site).}
\label{tab:tabley}
\end{figure}

\clearpage

\subsection*{Supplementary files}
Downloadable from web site: \url{http://www.lgm.upmc.fr/scolarietal/}

\end{document}